\documentclass[twocolumn,tighten]{aastex631}
\usepackage{amsmath,amstext}
\usepackage[T1]{fontenc}
\usepackage{natbib}
\usepackage{apjfonts} 
\usepackage{comment}
\usepackage{multirow}
\usepackage[figure,figure*]{hypcap}
\newcommand{\numextracted}{6048}
\newcommand{\linecatversion}{v4.1}
\newcommand{\clearer}{CLEAR$^{\mathrm{ER}}$}
\newcommand{\nmad}{$\sigma_{\mathrm{NMAD}}$}
\newcommand{\eazypy}{{\tt{eazy-Py}}}

\newcommand{\hst}{\textit{HST}}
\newcommand{\ngrst}{\textit{Roman}}
\newcommand{\jwst}{\textit{JWST}}

\newcommand{\lya}{Ly$\alpha$}

\newcommand{\hb}{\hbox{H$\beta$}}
\newcommand{\ha}{\hbox{H$\alpha$}}

\newcommand{\pab}{\hbox{Pa$\beta$}}
\newcommand{\oii}{\hbox{[O\,{\sc ii}]}}
\newcommand{\oiii}{\hbox{[O\,{\sc iii}]}}
\newcommand{\neiii}{\hbox{[Ne\,{\sc iii}]}}
\newcommand{\nii}{\hbox{[N\,{\sc ii}]}}
\newcommand{\mgii}{\hbox{Mg\,{\sc ii}}}
\newcommand{\sii}{\hbox{[S\,{\sc ii}]}}
\newcommand{\siii}{\hbox{[S\,{\sc iii}]}}

\newcommand{\todo}[1]{{\tt #1}}

\graphicspath{{./}{figures/}}

\begin{document}

\title{\large \textbf{CLEAR: Survey Overview, Data Analysis and Products}}

\correspondingauthor{Raymond C. Simons}
\email{raymond.simons@uconn.edu}

\author[0000-0002-6386-7299]{Raymond C. Simons}
\affiliation{Department of Physics, University of Connecticut, 196A Auditorium Road Unit 3046, Storrs, CT 06269 USA}

\author[0000-0001-7503-8482]{Casey Papovich}
\affiliation{Department of Physics and Astronomy, Texas A\&M University, College
Station, TX, 77843-4242 USA}
\affiliation{George P.\ and Cynthia Woods Mitchell Institute for
  Fundamental Physics and Astronomy, Texas A\&M University, College
  Station, TX, 77843-4242 USA}

\author[0000-0003-1665-2073]{Ivelina G. Momcheva}
\affil{Max-Planck-Institut für Astronomie, Königstuhl 17, D-69117 Heidelberg, Germany}

\author[0000-0003-2680-005X]{Gabriel Brammer}
\affil{Cosmic Dawn Centre, University of Copenhagen, Blegdamsvej 17, 2100 Copenhagen, Denmark}

\author[0000-0001-8489-2349]{Vicente Estrada-Carpenter}
\affiliation{Department of Astronomy \& Physics, Saint Mary's University, 923 Robie Street, Halifax, NS, B3H 3C3, Canada}

\author[0000-0001-8519-1130]{Steven L. Finkelstein}
\affil{Department of Astronomy, The University of Texas at Austin, Austin, TX, 78759 USA}

\author[0000-0001-8023-6002]{Catherine M. Gosmeyer}
\affiliation{Kairos Aerospace, Sunnyvale, CA  94085, USA}

\author[0000-0002-7547-3385]{Jasleen Matharu}
\affiliation{Cosmic Dawn Center, Niels Bohr Institute, University of Copenhagen, R\aa dmandsgade 62, 2200 Copenhagen, Denmark\\}

\author[0000-0002-1410-0470]{Jonathan R. Trump}
\affil{Department of Physics, University of Connecticut, 196A Auditorium Road Unit 3046, Storrs, CT 06269 USA}

\author[0000-0001-8534-7502]{Bren E. Backhaus }
\affil{Department of Physics, University of Connecticut, 196A Auditorium Road Unit 3046, Storrs, CT 06269 USA}

\author[0000-0001-8551-071X]{Yingjie Cheng}
\affiliation{University of Massachusetts Amherst, Amherst, MA, 01003, USA}

\author[0000-0001-7151-009X]{Nikko J. Cleri}
\affiliation{Department of Physics and Astronomy, Texas A\&M University, College Station, TX, 77843-4242 USA}
\affiliation{George P.\ and Cynthia Woods Mitchell Institute for Fundamental Physics and Astronomy, Texas A\&M University, College Station, TX, 77843-4242 USA}

\author[0000-0001-7113-2738]{Henry C. Ferguson}
\affiliation{Space Telescope Science Institute, 3700 San Martin Drive, Baltimore, MD, 21218 USA}

\author[0000-0002-0496-1656]{Kristian Finlator}
\affiliation{Cosmic Dawn Center, Niels Bohr Institute, University of Copenhagen, Jagtvej 128, 2200 Copenhagen N, Denmark\\}
\affil{Department of Astronomy, New Mexico State University, Las Cruces, NM 88003, USA\\}

\author[0000-0002-7831-8751]{Mauro Giavalisco}
\affil{Astronomy Department, University of Massachusetts, Amherst, MA, 01003 USA} 

\author[0000-0001-7673-2257]{Zhiyuan Ji}
\affil{Astronomy Department, University of Massachusetts, Amherst, MA, 01003 USA} 

\author[0000-0003-1187-4240]{Intae Jung}
\affiliation{Space Telescope Science Institute, Baltimore, MD, 21218, USA}

\author[0000-0003-3130-5643]{Jennifer M. Lotz}
\affiliation{Gemini Observatory/NSF's National Optical-Infrared Astronomy Research Laboratory, 950 N. Cherry Ave., Tucson, AZ 85719, USA}

\author[0000-0003-3351-0878]{Rosalia O'Brien}
\affiliation{Department of Physics and Astronomy, Texas A\&M University, College Station, TX, 77843-4242 USA}
\affiliation{School of Earth and Space Exploration, Arizona State University, Tempe, AZ 85287-1404, USA}

\author[0000-0001-7393-3336]{Rosalind E. Skelton}
\affiliation{South African Astronomical Observatory, P.O. Box 9, Observatory, Cape Town 7935, South Africa}

\author[0000-0001-8514-7105]{Vithal Tilvi}
\affiliation{School of Earth and Space Exploration, Arizona State University, Tempe, AZ 85287, USA}

\author[0000-0001-6065-7483]{Benjamin Weiner}
\affil{MMT/Steward Observatory, 933 N. Cherry St., University of Arizona, Tucson,
AZ 85721, USA}


\begin{abstract} 
We present an overview of the CANDELS Lyman-$\alpha$ Emission At Reionization (CLEAR) survey. CLEAR is a 130 orbit program of the \textit{Hubble Space Telescope} using the Wide Field Camera 3 (WFC3) IR G102 grism.   CLEAR targets 12 pointings divided between the GOODS-N and GOODS-S fields of the Cosmic Assembly Near-IR Deep Extragalactic Legacy Survey (CANDELS).  Combined with existing spectroscopic data from other programs, the full CLEAR dataset includes spectroscopic imaging of these fields over 0.8--1.7~\micron.  In this \textit{Paper}, we describe the CLEAR survey, the survey strategy, the data acquisition, reduction, processing, and science products and catalogs released alongside this paper.  The catalogs include emission line fluxes and redshifts derived from the combination of the photometry and grism spectroscopy for \numextracted{} galaxies, primarily ranging from $0.2 \lesssim z \lesssim 3$. We also provide an overview of CLEAR science goals and results.  In conjunction with this \textit{Paper} we provide links to electronic versions of the data products, including 1D + 2D extracted spectra and emission line maps.
\end{abstract}


\section{Introduction}
The spectroscopic capabilities of the \textit{Hubble Space Telescope}
(\hst) provide a novel method to characterize and study the
evolution of galaxies.   Lying above the Earth's atmosphere, \hst\ is able to produce high-angular resolution images without the high sky backgrounds that plague ground-based observations.  Slitless
spectroscopy from \hst\ therefore has two main advantages compared to
terrestrial observations:  it provides the spatial quality of \hst\
($0\farcs1-0\farcs2$ FWHM) with low backgrounds.   Since the
installation of the Wide Field Camera 3 (WFC3), we have seen a revolution in the slitless spectroscopy of distant galaxies.  Primarily this has been provided by the grisms in the WFC3 IR camera, G102 and G141, which disperse light from 0.8--1.1~\micron, and 1.1--1.7~\micron, respectively, with low spectral resolution ($R = \Delta \lambda / \lambda \sim 200$ and $\sim 100$, respectively).  From initial work with the \textit{Early Release Science} (ERS) programs \citep{vandokkum10,straughn11}, the community has carried out a series of programs including both targeted deep and wide-field surveys (e.g., FIGS, \citealt{pirzkal17}; 3D-HST, \citealt{momcheva16}; GLASS, \citealt{treu15}; AGHAST, \citealt{weiner12}; ; MAMMOTH-Grism \citealt{wang22}; 3D-DASH \citealt{mowla22}; MUDF \citealt{revalski23}), snapshot programs (e.g., WISPS, \citealt{atek10}), and targeted observations of transient sources (such as SNe, e.g., \citealt{rodney12}).  

Following in the legacy of these studies, we present here the dataset from the CANDELS Lyman-$\alpha$ (\lya) Emission at Reionization (CLEAR) survey.  CLEAR is a \hst\ Cycle 23 program that obtained deep (10 to 12-orbit depth) observations with the \hst/WFC3 using the G102 grism in the IR camera. The observations (130 orbits total) cover 12 fields in the GOODS-N and GOODS-S fields overlapping the WFC3 imaging footprint of the Cosmic Assembly Near-IR Deep Extragalactic Legacy Survey (CANDELS; \citealt{2011ApJS..197...35G, 2011ApJS..197...36K}).  The primary goal of CLEAR was to characterize the evolution of the Lyman-$\alpha$ equivalent width distribution at $6 < z < 8$ and to interpret this in the context of reionization---as the IGM of the Universe transitions from one that is mostly ionized at $z < 6$ to one that is mostly neutral at $z > 6$ \citep{robertson13}.  This is important as \lya\ emission is sensitive to neutral \ion{H}{1} fractions of 0.01 to 1.0 \citep{mcquinn07}, and there is a need to trace \lya\ from the ionized universe at $z =6-6.5$ to the neutral universe at $z > 7$ with systematic, homogeneous surveys.  In addition, the CLEAR pointings overlap with G102 and G141 observations from a number of previous programs (including the FIGS, AGHAST, and 3DHST surveys). Together with CLEAR, this dataset provides  slitless spectroscopy at the spatial resolution of \hst\ covering most of the $Y$, $J$, and $H$ bands, $0.8-1.7$~\micron. This enables a wide range of science using strong emission lines and stellar continuum features in the rest-frame optical, that are redshifted into the near-IR and observable in the grism data. Furthermore, a major advantage of slitless spectroscopy is that it provides a spectrum for all galaxies in the field---target preselection is {\emph{not}} required.

Here, we describe the CLEAR survey strategy, data acquisition, reduction, and science products. Along with this paper, we release the high-level 1D and 2D spectra, emission line maps, and redshift/line catalogs produced through this survey. 

To date, the CLEAR dataset has been used to study the evolution of: the \lya\ equivalent-width distribution into the epoch of reionization
\citep{jung22}, galaxy stellar population properties including
ages, star-formation histories, and chemical enrichment histories \citep{estrada-carpenter19,estrada-carpenter20}, emission-line ratios, metallicities and ionization properties in galaxies in both a spatially-integrated \citep{backhaus22a, papovich22} and spatially-resolved sense \citep{simons21, matharu22, backhaus22b}, supermassive black-holes \citep{yang21}, \pab{} as a star-formation indicator \citep{cleri22a}, high-ionization [NeV] emission in galaxies (\citealt{cleri22b,cleri2023}), and the mass-metallicity relation \citep{henry21, papovich22}. These studies demonstrate that the CLEAR data products provide a resource for identifying and characterizing the properties of galaxies over a wide range of redshift, including the peak of the cosmic star-formation density \citep{madau14} and
supermassive black-hole accretion density \citep{brandt15}.

The outline for this paper is as follows.  In
Section~\ref{section:design} we describe the design of the survey, and provide the details of the CLEAR observing program.  In Section~\ref{section:ancillary} we describe the ancillary \hst\ grism datasets that we include in our analysis of the CLEAR dataset.  In Section~\ref{section:photometry} we describe the multiwavelength photometric catalog we employ for analysis of the CLEAR galaxies.  In Section~\ref{section:reduction} we describe the process for data reduction, calibration, spectral extractions, and derived quantities including redshifts and emission line fluxes from the grism spectroscopy. In Section ~\ref{section:catalogs}, we discuss the catalogs and data products released alongside this paper. In Section~\ref{section:science} we discuss the CLEAR science, and provide additional examples of using the data for science.  Finally, in Section~\ref{section:summary} we provide a brief summary.   Throughout this paper, we use magnitudes on the \textit{Absolute Bolometric} system \citep{oke83} and a cosmology that assumes $\Omega_{m,0}=0.3$, $\Omega_{\Lambda,0}=0.7$, and $H_0 = 70$~km s$^{-1}$ Mpc$^{-1}$.    We use a Chabrier-like IMF for any quantities such as stellar masses and star-formation rates (SFR)s. 


\section{Survey Design and Data Acquisition}\label{section:design}
\begin{figure*}[h!]
\epsscale{1.15}
\plotone{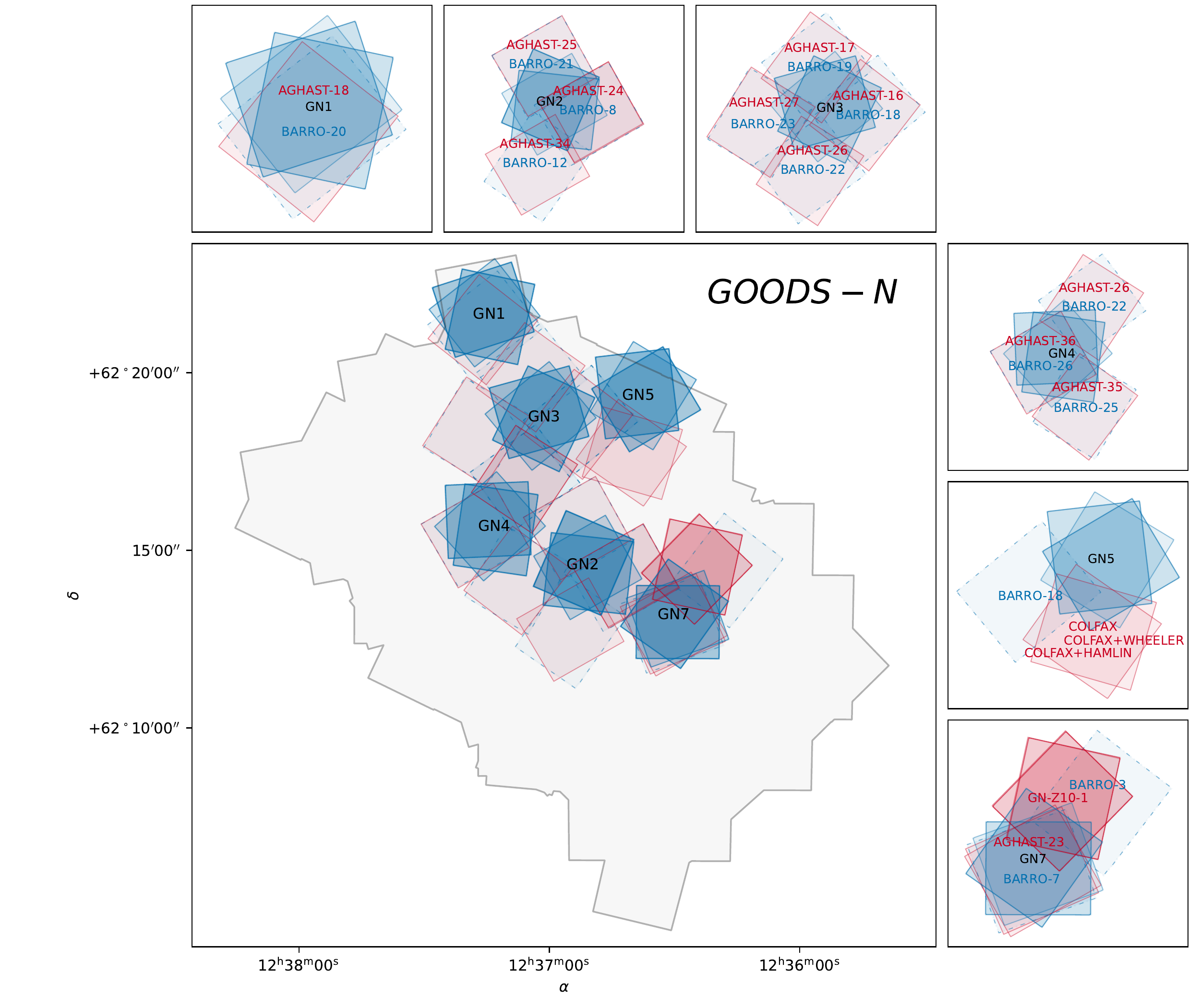}
\caption{Footprints of the CLEAR fields in GOODS-North. Main panel: CLEAR G102 observations (blue, solid line) overlaid on the footprint of the CANDELS WFC3 imaging (gray). Also shown are the overlapping pointings of G102 (blue, dashed line) and G141 (red, solid line) grism observations from ancillary programs that are included in our data reduction and analysis. Top and bottom side panels: zoomed-in view of each of the six CLEAR fields. \label{fig:goodsn_clear_footprints}}
\epsscale{1.0}
\end{figure*}

\begin{figure*}[h!]
\epsscale{1.15}
\plotone{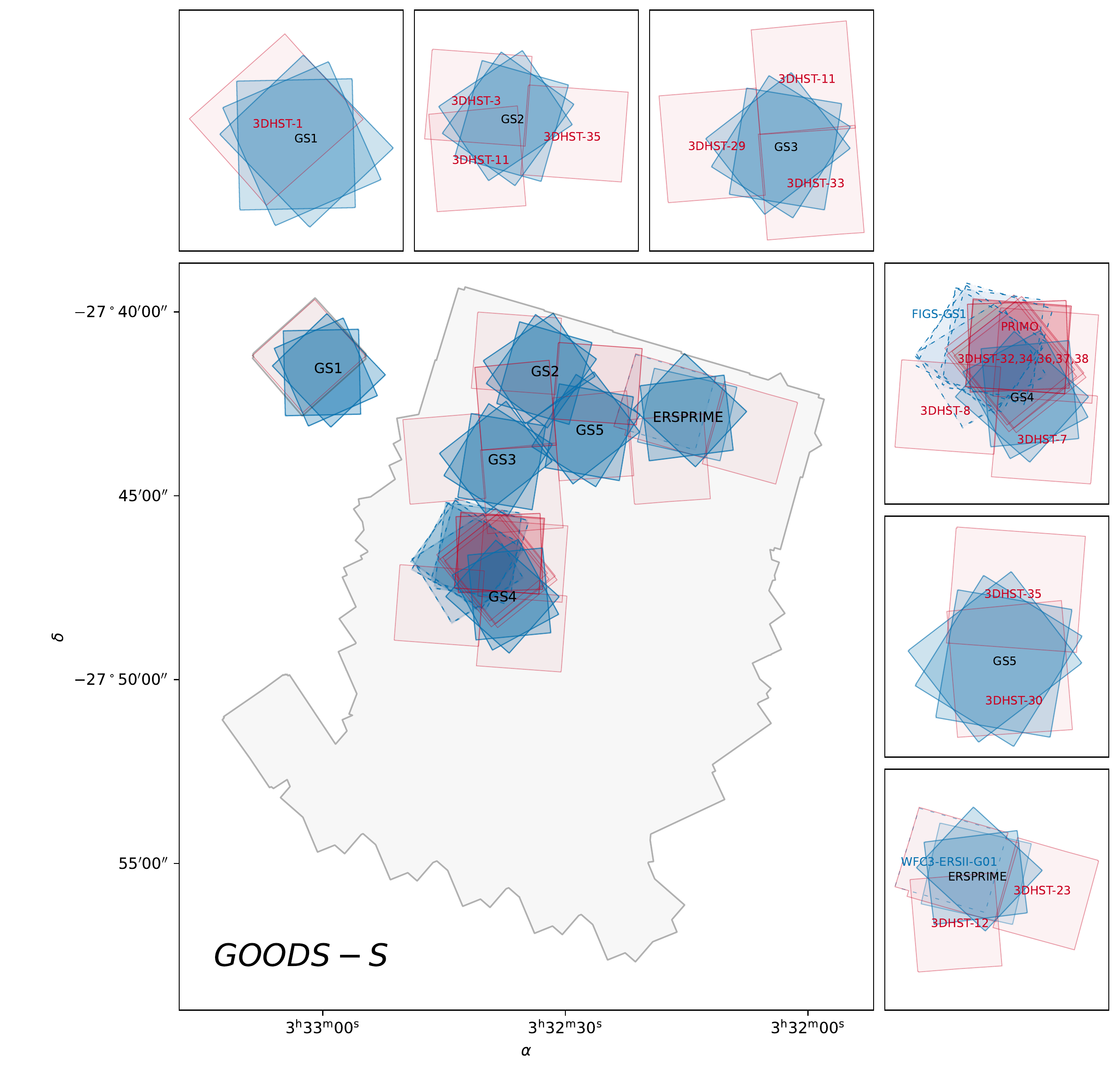}
\caption{Footprints of the CLEAR fields in GOODS-South. Same as Figure \ref{fig:goodsn_clear_footprints}. Of note, the GS4 field overlaps with the Hubble Ultra Deep Field which includes ancillary 8-orbit depth G141 observations from the 3D-HST survey \citep{momcheva16} and 40-orbit depth G102 observations from the FIGS survey \citep{pirzkal17}.\label{fig:goodss_clear_footprints}}
\epsscale{1.0}
\end{figure*}

The CLEAR program was designed in area and depth to survey a sufficient number of high-redshift galaxies to the line flux sensitivities needed to achieve the primary science goals of the survey---constraints on the \lya\ line emission in $6 < z < 8$ galaxies to limits of $\simeq10^{-17}$~erg s$^{-1}$ cm$^{-2}$.  We targeted 12 new fields with WFC3, evenly divided between the GOODS-N and GOODS-S galaxy fields. Figures~\ref{fig:goodsn_clear_footprints} and \ref{fig:goodss_clear_footprints} show the locations of the CLEAR pointings.

\begin{deluxetable*}{lcclccc}
\tablecolumns{7}
\tablewidth{0pt}
\tablecaption{WFC3 Observing Summary of CLEAR Fields \label{table:hstobs}}
\tablehead{
\colhead{Field Name} & 
\colhead{R.A. (J2000 ICRS)} & 
\colhead{Decl. (J2000 ICRS)} & 
\colhead{Observing Date(s)} & 
\colhead{\# of Orbits} & 
\colhead{ORIENT (deg)} & 
\colhead{Observing Sequence}
}
\startdata
\multirow{3}{*}{GN1} & \multirow{3}{*}{12 37 13.54} & \multirow{3}{*}{$+$62 21 50.4} &  2016 Aug 30 & 4 & $-$102 & F105W, G102, G102 \\
  &  &  &  2016 Oct 07 &  2 & $-$142 & G102, G102, F105W \\
  &  &  &  2016 Oct 26-27, 29 & 4 & $-$162 & F105W, G102, G102 \\\hline
\multirow{4}{*}{GN2} & \multirow{4}{*}{12 36 52.93} & \multirow{4}{*}{$+$62 14 24.4} & 2016 Mar 08 & 2 & $+$66.7 & G102, G102, F105W  \\
  & & & 2016 Apr 05 & 2 & $+$29.7 & F105W, G102, G102 \\
  & & & 2016 May 01 & 4 & $-$6.4 & G102, G102, F105W \\
  & & & 2017 Feb 19 & 2 & $+$66.7 & G102, G102, F105W \\\hline
\multirow{3}{*}{GN3} & \multirow{3}{*}{12 37 00.16} & \multirow{3}{*}{$+$62 18 56.4} & 2016 Sep 11 & 4 & $-$115.61 & F105W, G102, G102  \\
  & & & 2016 Oct 07 & 2 & $-140.5$ & G102, G102, F105W \\
  & & & 2016 Oct 30-31 & 4 & $-164.5$ & F105W, G102, G102 \\ \hline
\multirow{3}{*}{GN4} & \multirow{3}{*}{12 37 12.34} & \multirow{3}{*}{$+$62 15 50.4} & 2015 Nov 15 & 4 & $-177.6$ & F105W, G102, G102 \\
  & & & 2016 Aug 25, 28 & 4 & $-98.3$ & F105W, G102, G102 \\
  & & & 2016 Oct 01 & 2 & $-138.3$ & G102, G102, F105W \\ \hline
\multirow{3}{*}{GN5} & \multirow{3}{*}{12 36 38.91} & \multirow{3}{*}{$+$62 19 08.4} & 2016 Feb 02 & 4 & $96.7$ & G102, G102, F105W \\
 & & & 2016 Mar 7 & 2 & 58.6 & G102, G102, F105W \\
 & & & 2016 Apr 09, 10 & 4 & 30.7 & F105W, G102, G102 \\ \hline
\multirow{3}{*}{GN7} & \multirow{3}{*}{12 36 31.54} & \multirow{3}{*}{+62 13 00.4}  & 2016 Mar 04, 05 & 4 & $54.7$ & G102, G102, F105W \\
 & & & 2016 Apr 05 & 2 & 19.7 & F105W, G102, G102 \\
 & & & 2016 Apr 26, 27 &  4 & $-$0.3 & G102, G102, F105W \\ \hline
 \multirow{3}{*}{GS1} & \multirow{3}{*}{03 32 32.40} & \multirow{3}{*}{$-$27 42 22.4} & 
 2015 Dec 20, 2016 Jan 22 & 4 & $-$88.7 & F105W, G102, G102 \\ 
  & & & 2016 Jan 22 & 4 & $-66.3$ & F105W, G102, G102 \\
 & & & 2016 Jan 28 & 4 & $-46.3$ & F105W, G102, G102 \\ \hline
 \multirow{3}{*}{GS2} & \multirow{3}{*}{03 32 32.16} & \multirow{3}{*}{$-$27 41 24.5} & 
 2015 Nov 14, 16 & 4 & $-$146.2 & G102, G102, F105W \\ 
 & & & 2015 Nov 30, Dec 07 & 4 & $-125.9$ & F105W, G102, G102 \\
 & & & 2015 Dec 15, 16 & 4 & $-$105.9 & F105W, G102, G102 \\ \hline
 \multirow{3}{*}{GS3} & \multirow{3}{*}{03 32 37.57} & \multirow{3}{*}{$-$27 43 48.53} & 2015 Nov 18, 19 & 4 & $-$142.3 & G102, G102, F105W \\
 & & & 2015 Dec 13, 17 & 4 & $-$122.3 & F105W, G102, G102 \\
 & & & 2016 Jan 04, 07 & 4 & $-$99.3 & F105W, G102, G102 \\ \hline
 \multirow{3}{*}{GS4} & \multirow{3}{*}{03 32 36.94} & \multirow{3}{*}{$-$27 47 56.5}  & 2016 Aug 06, 07 & 4 & $95.4$ & G102, G102, F105W \\ 
 & & & 2016 Aug 11, 18 & 4 & 138.3 & G102, G102, F105W \\
 & & & 2016 Aug 15, 19 & 4 & 118.3 & G102, G102, F105W \\ \hline
 \multirow{3}{*}{GS5} & \multirow{3}{*}{03 32 26.73} & \multirow{3}{*}{$-$27 43 00.5} & 2015 Nov 16, 17 & 4 & $-$142.5 & G102, G102, F105W \\ 
 & & & 2015 Dec 11, 13 & 4 & $-$121.8 & F105W, G102, G102 \\
 & & & 2015 Dec 19 & 4 & $-$99.8 & F105W, G102, G102 \\ \hline
 \multirow{3}{*}{ERSPRIME} & \multirow{3}{*}{03 32 13.78} & \multirow{3}{*}{$-$27 42 52.5} & 2016 Aug 15 & 4 & 137.0 & G102, G102, F105W \\
 & & & 2016 Sep 24 & 2 & 167.0 & G102, G102, F105W \\
 & & & 2016 Oct 6 & 4 & $-173.0$ & G102, G102, F105W \\ \hline\hline
& & &  \multicolumn{1}{r}{Total: }  & 130 & & \\
  \enddata
\end{deluxetable*}


\subsection{Target Field Selection}

The primary goal of the CLEAR survey was to constrain the amount of \lya\ emission from galaxies in the Epoch of Reionization. To that end, we selected fields in GOODS-N and -S which maximized the number of photometrically-selected target galaxies over the redshift range $6 < z < 8$.

To select the fields for CLEAR, we used the LBG catalog of \citet{Finkelstein15}.  This provided $>$6 potential pointings in GOODS-N and GOODS-S each. We then downselected to 6 in each field.  The CLEAR fields are illustrated in Figures~\ref{fig:goodsn_clear_footprints} and \ref{fig:goodss_clear_footprints}. They are labeled "GN1--GN5, GN7" in GOODS-N (where they are non-sequential as we dropped a GN6 field) and "GS1--GS5" in GOODS-S.  GS1 overlaps with the HUDF ACS parallel field \citep{Beckwith06} and the sixth field in GOODS-S coincides mostly with the WFC3/ERS field \citep{straughn11} which we designate ``ERSPRIME''. The coordinates of the fields, including the number of new orbits provided by CLEAR, are given in Table~\ref{table:hstobs}.  The field area of CLEAR is significantly larger than the typical spatial extent of ionized structures during the epoch of reionization (e.g., \citealt{Ocvirk20}). Moreover, cosmic variance is not an issue for CLEAR as the GOODS-N and -S fields are sufficiently separated on the sky, and the redshift range $6 < z < 8.2$ over which the G102 wavelength coverage is sensitive to redshifted \lya\ provides sufficient volume for galaxy populations to be unrelated in redshift.

\subsection{Considerations for the \textit{Hubble Space Telescope} Observations}

We split each orbit of the \hst/WFC3 observations into a direct image (F105W) and two G102 grism exposures of the same pointing.  Each WFC3 exposure used the \texttt{MULTIACCUM} mode, with the sample sequencing (\texttt{SAMP-SEQ}) and number of samples (\texttt{NSAMP}) depending on the type of observation.  Each WFC3/F105W direct image comprises a single iteration (exposure) with \texttt{SPARS25} and \texttt{NSAMP} = 11. This produced 303~s observations.  The G102 exposures used a single iteration with \texttt{SPARS100} and either \texttt{NSAMP}=12 or 13 samples---depending on the amount of usable time per orbit. This provided a total exposure time of 1103 or 1203~s per exposure.    In all cases, we adopted the dither pattern employed by 3D-HST \citep{momcheva16} to match the sampling of those data as closely as possible. 

We observed each pointing in CLEAR using two orbits at a single position angle (ORIENT), repeating the pattern above.  We required additional orbits to have a position angle offset by at least $20^\circ$.  That requirement ensures that the spectral trace from each object falls on different portions of the detector and that contamination from nearby sources occurs in only a single PA \citep[see, e.g., the discussion in][]{estr19}.  Table~\ref{table:hstobs} lists the ORIENTs and number of orbits per pointing.  

In addition, WFC3 $Y$-band exposures are known to suffer time-variable backgrounds during the \hst\ orbit \citep{lotz17}.  The origin of this background is due to \ion{He}{1} 10830~\AA\ emission from the Earth's atmosphere when \hst\ observes at low limb angles.  This background is strongest when \hst\ is not in the Earth's shadow, which occurs at the start or end of each orbit.  Following \citeauthor{lotz17} we predicted the \hst\ ephemeris for each of our orbits and scheduled the sequence of F105W direct images and two G102 grism exposures so that the latter were taken when \hst\ was in the shadow of the Earth. In doing so, the grism observations were protected from the \ion{He}{1} background.  As a tradeoff, the F105W imaging suffers from higher backgrounds. This was acceptable as those images are used only for alignment while the grism spectroscopy is required for the primary science.   Table~\ref{table:hstobs} lists the observing sequence of F105W and G102 during the observation where either the direct image occurs first in the orbit (F105W, G102, G102) or last in the orbit (G102, G102, F105W).


\section{Ancillary Observations}\label{section:ancillary}

\subsection{Imaging data}\label{subsection:imaging}

The CLEAR pointings lie in the well-studied GOODS-S and GOODS-N galaxy fields. These fields have extensive UV to IR imaging. We refer the reader to Table 3 of \citet{skelton14} for full details, and briefly describe the relevant imaging datasets here.

\emph{HST}/ACS\,+\,WFC3 imaging is available in 7 and 10 bands in GOODS-N and GOODS-S, respectively. The majority of this \emph{HST} imaging is provided by three large programs: the Great Observatories Origins Deep Survey (GOODS; \citealt{2004ApJ...600L..93G}), the CANDELS Multi-Cycle Treasury Project \citep{2011ApJS..197...35G, 2011ApJS..197...36K} and the 3D-HST Treasury Program \citep{momcheva16, 2012ApJS..200...13B, skelton14}.

In addition, UV to 8 $\mu$m imaging is available from a number of ground- and space-based observatories: KPNO 4 m/Mosaic (\emph{U}; \citealt{capak04}), VLT/VIMOS (\emph{U-R}; \citealt{nonino09}), WFI 2.2m (\emph{U38-B-V-R$_C$-I}; \citealt{hildebrandt06, erben05}), Keck/LRIS (\emph{G-R$_S$}; \citealt{steidel03}), Subaru/Suprime-Cam (\emph{B-V-R$_C$-I$_C$-z$^\prime$} and 14 medium bands; \citealt{capak04, cardamone10}), Subaru/MOIRCS (\emph{J-H-K$_S$}; \citealt{kajisawa11}), VLT/ISAAC (\emph{J-H-K$_S$}; \citealt{retzlaff10, wuyts08}), CFHT/WIRCam (\emph{J-K$_S$}; \citealt{hsieh12}), and {\emph{Spitzer}}/IRAC (3.6-4.5-5.8-8 $\mu$m; \citealt{ashby13, dickinson03}).

\subsection{Grism data}\label{section:grism_data} 

To supplement the CLEAR G102 grism spectroscopy, we queried the Mikulski Archive for Space Telescopes (MAST) for G102 (0.8 $\mu$m - 1.1 $\mu$m) and G141 (1.1 $\mu$m - 1.7 $\mu$m) observations that overlap the CLEAR footprint. We retrieved a total of 52 orbits of G102 and 76 orbits of G141 observations---taken through the programs listed in Table \ref{table:ancillarydata}. 

We refer to this combined dataset as `\clearer' for CLEAR-Extended Release. The distribution of G102 and G141 exposure times for the objects extracted as a part of the full \clearer dataset are shown in Figure \ref{fig:exposure_time}. Of note, \clearer{} includes ultra-deep 40-orbit G102 spectra in the Hubble Ultra Deep-Field (the `GS4' pointing of CLEAR) taken as a part of the FIGS program \citep{pirzkal17}. The FIGS field contributes to the high-depth G102 tail of the dataset (see the right panel of Figure \ref{fig:exposure_time}). 

Combined, the G102 and G141 grisms cover a continuous wavelength range of 0.8 to 1.7 $\mu$m. The visibility windows of bright rest UV - NIR lines are shown for both grisms in Figure \ref{fig:lines_redshift}. With joint grism coverage, we are able to capture a more complete set of emission lines for the same galaxy. As an example, with both grisms employed, the full R$_{23}$ complex (H$\beta$, \oiii, \oii) is visible in galaxies over a redshift range of 1.2 $<$ z $<$ 2.4. With only one of the grisms, this range is considerably smaller---1.2 $<$ z $<$ 1.3 for the G102 grism alone and 2.0 $<$ z $<$ 2.4 for the G141 grism alone.

\begin{deluxetable*}{lccccccc}
\tablecolumns{7}
\tablewidth{0pt}
\tablecaption{Ancillary WFC3 Grism Observations Overlapping the CLEAR Pointings}\label{table:ancillarydata}
\tablehead{
\colhead{} & 
\multicolumn{2}{c}{Number of Orbits\tablenotemark{a}} & 
\colhead{\hst} & 
\colhead{Proposal} & 
\colhead{Principal} & 
\colhead{Survey or } \\
\colhead{Field} & 
\colhead{G102} & 
\colhead{G141} & 
\colhead{Cycle} & 
\colhead{ID} & 
\colhead{Investigator} & 
\colhead{Pointing Name} 
}
\startdata
\multirow{4}{*}{GOODS-N} &   \nodata &  21      & 17 & 11600 &  Weiner   &  AGHAST  \\
                         &   \nodata &  3       & 19 & 12461 & Riess     &  SN COLFAX \\
                         &   22      & \nodata  & 21 & 13420 & Barro     &            \\
                         &   \nodata & 9        & 25 & 13871 & Oesch     &            \\ \hline
\multirow{4}{*}{GOODS-S} &   1       & 1        & 17 & 11359 & O'Connell & WFC3/ERS  \\
                         &   \nodata & 13       & 18 & 12099 & Riess     & GEORGE, PRIMO\\
                         &   \nodata & 29       & 18 & 12177 & van Dokkum & 3D-HST \\ 
                         &   29      & \nodata  & 22 & 13779 & Malhotra &  FIGS \\ \hline
Total                    &  52     &  76   &    &  &  & \\                     
        \enddata
        \tablenotetext{a}{The number of orbits listed is of the subset of pointings overlapping the CLEAR field, and does not reflect the total number of orbits of the respective programs.}
        \label{tab:2}

\end{deluxetable*}


\section{Updated 3D-HST Photometric Catalogs}\label{section:photometry}

As a part of the 3D-HST survey, \citet{skelton14} carried out source detection and photometric analysis on the full set of imaging described in \S\ref{subsection:imaging}. The resulting photometric catalogs are available on the 3D-HST website\footnote{\url{https://archive.stsci.edu/prepds/3d-hst/}} (`v4.1' as of this publication). These are the root catalogs used for the \clearer{} dataset. As described above and in Table \ref{table:hstobs}, we supplement this catalog with {\emph{HST}}/WFC3 F105W photometry for the sources in the CLEAR footprint. The F105W fluxes and uncertainties are measured in a manner that is consistent with \citet{skelton14}. We also incorporate new ground-based spectroscopic redshifts (`z\_spec') from the KMOS-3D \citep{2019ApJ...886..124W} and MOSDEF surveys \citep{2015ApJS..218...15K} in GOODS-S and GOODS-N. The original compilation of spectrosopic redshifts in the 3D-HST catalog derives from the MOIRCS Deep Survey catalog in GOODS-N \citep{2011PASJ...63S.379K} and the FIREWORKS catalog in GOODS-S \citep{2008ApJ...682..985W}---see \citet{skelton14} for details. We supplant these redshifts with those from the KMOS-3D (quality flag = 1 in their catalog; N = 43) and MOSDEF (quality flag $\ge3$ in their catalog; N = 143) surveys, when the latter two are available.

With these updates to the catalog, we use the \eazypy{}\footnote{\url{https://github.com/gbrammer/eazy-py}} code (a Python photometric analysis and redshift tool based on {\tt{EAZY}}; \citealt{brammer08}) to derive new zeropoint corrections, photometric redshifts, and rest-frame colors for the full 3D-HST sample in GOODS-S and GOODS-N. We also use {\tt{eazy-py}} to derive new broadband-based estimates of stellar masses, star-formation rates, and dust attenuation A$_V$.

We adopt the set of `fsps\_QSF\_12\_v3'  Flexible Stellar Population Synthesis continuum templates (FSPS; \citealt{2009ApJ...699..486C, 2010ApJ...712..833C}) available in the \eazypy{} library. The FSPS templates assume a \citet{2003PASP..115..763C} initial mass function and were constructed to span a range of galaxy types (following the methodology of \citealt{2007AJ....133..734B, brammer08}). The updated version of the 3D-HST photometric catalog (`v4.6'\footnote{The public photometric catalog jumps from version v4.1 to v4.6, skipping over intermediate internal CLEAR team releases.}) is released alongside this paper. The full \eazypy{} parameter file that is used in the run is also provided in the release. The columns of the catalog are described in Table 10 of \citet{skelton14}, with two new columns of F105W flux and flux uncertainties provided by CLEAR. In addition to the photometric catalog, we also release a catalog of \eazypy{}-derived galaxy properties. The contents of this catalog are described in Table \ref{table:eazy_catalog}.

\section{Data Reduction and Processing}\label{section:reduction}

We process the complete dataset of grism and imaging observations described in \S\ref{section:design}, \S\ref{section:ancillary} and Tables \ref{table:hstobs} and \ref{table:ancillarydata} using the grism redshift and line analysis software {\tt{Grizli}}\footnote{\url{https://github.com/gbrammer/grizli}} \citep{2019ascl.soft05001B}. As described below, {\tt Grizli} performs end-to-end processing of {\emph{HST}} imaging and slitless spectroscopy datasets. This includes retrieving and pre-processing the raw observations, performing astrometric alignment, modeling contamination from overlapping spectra, extracting the 1D and 2D spectra of individual sources, fitting continuum + emission-line models, and generating emission-line maps.

\subsection{Pre-processing} 

We use {\tt{Grizli}} to retrieve the observations described in Table \ref{table:hstobs} and \ref{tab:2} from the \textit{Barbara A. Mikulski Archive for Space Telescopes} (MAST) archive. Then, the raw observations are reprocessed with the \texttt{calwf3} pipeline and corrections for variable sky backgrounds \citep{2016wfc..rept...16B} are applied. Cosmic rays and hot pixels are identified with the AstroDrizzle software \citep{2012drzp.book.....G}.  Flat field corrections are applied to the G102 (G141) grism exposures using the F105W (F140W) calibration images. We use the ``Master Sky'' constructed in \citet{2015wfc..rept...17B} to carry out sky subtraction.  Using the deeper 3D-HST {\emph{HST}}/WFC3 F140W galaxy catalog of these fields \citep{skelton14} as reference, a relative astrometric correction is applied to the data. 




\begin{figure}
\includegraphics[width=\columnwidth]{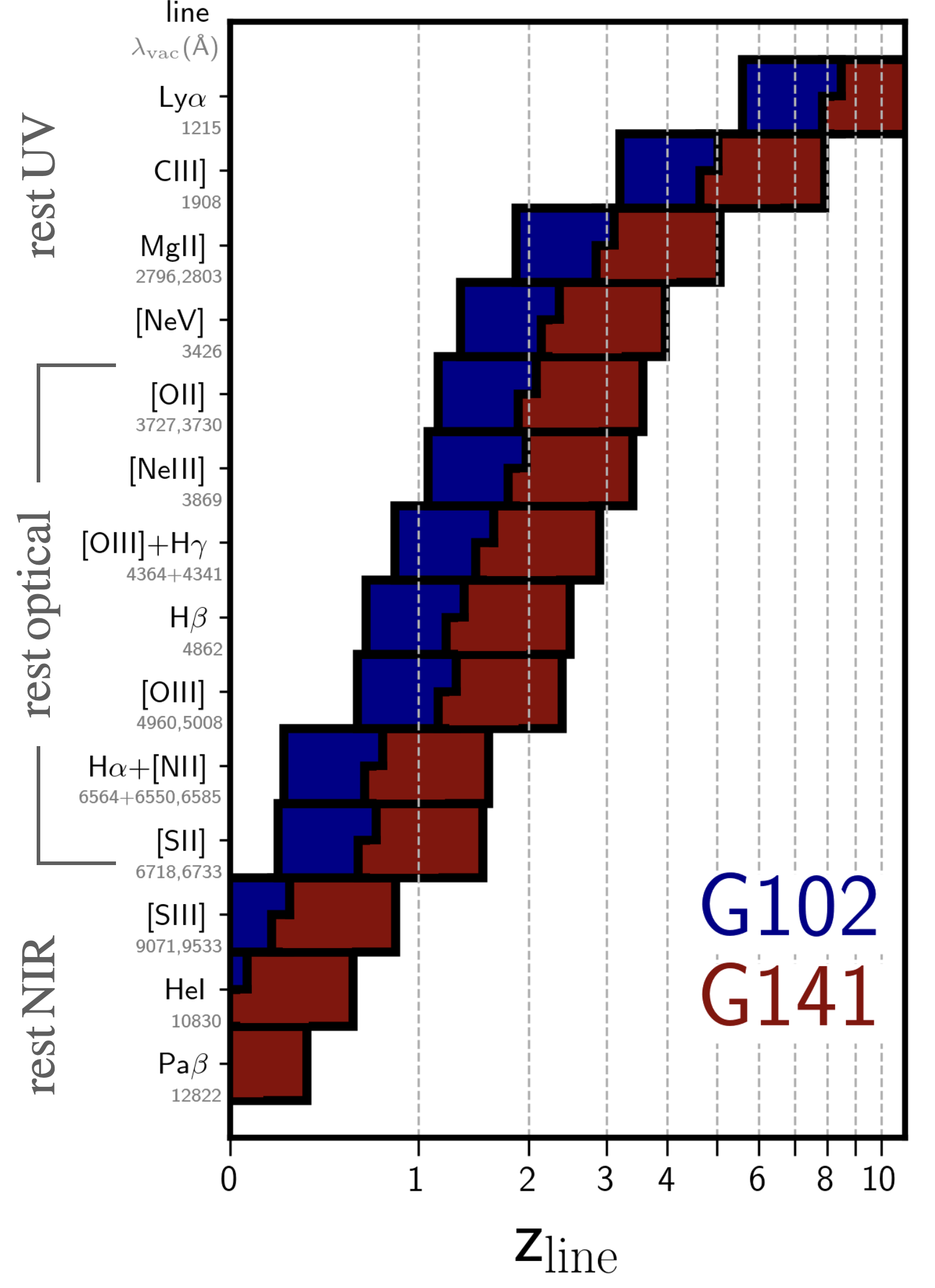}
\caption{The redshift ranges over which different emission lines are observable with the {\emph{HST}}/WFC3 G102 and G141 grisms are shown.  Each row indicates the redshifts of the G102 (blue) and G141 (red) grism spectroscopic coverage. The emission lines are labeled along the ordinate.}\label{fig:lines_redshift}
\end{figure}

\begin{figure*}
\centering
\includegraphics[width=\textwidth]{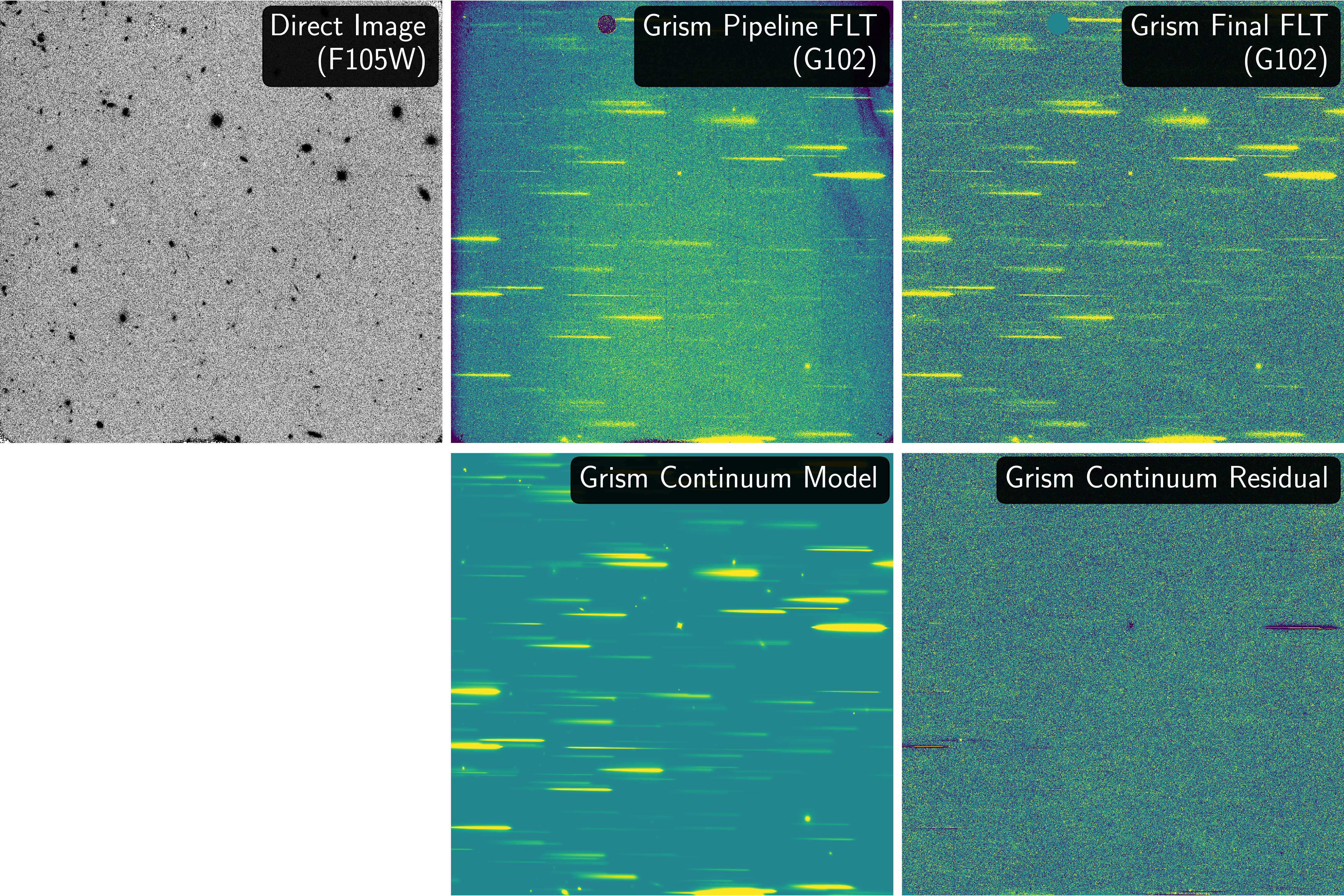}\label{fig:detector_model}
\caption{The reduction and continuum modeling of a single G102 grism exposure. The top left panel shows the direct \emph{HST}/WFC3 F105W image. The top middle and right panel show the pipeline FLT and the background- and flat-fielded processed final FLT. The wavelength increases from left to right. The bottom middle panel shows the continuum model for the sources in the field and the bottom right panel shows the residual of the observations and continuum model.}
\end{figure*}

\begin{figure*}
\centering
\includegraphics[width=\textwidth]{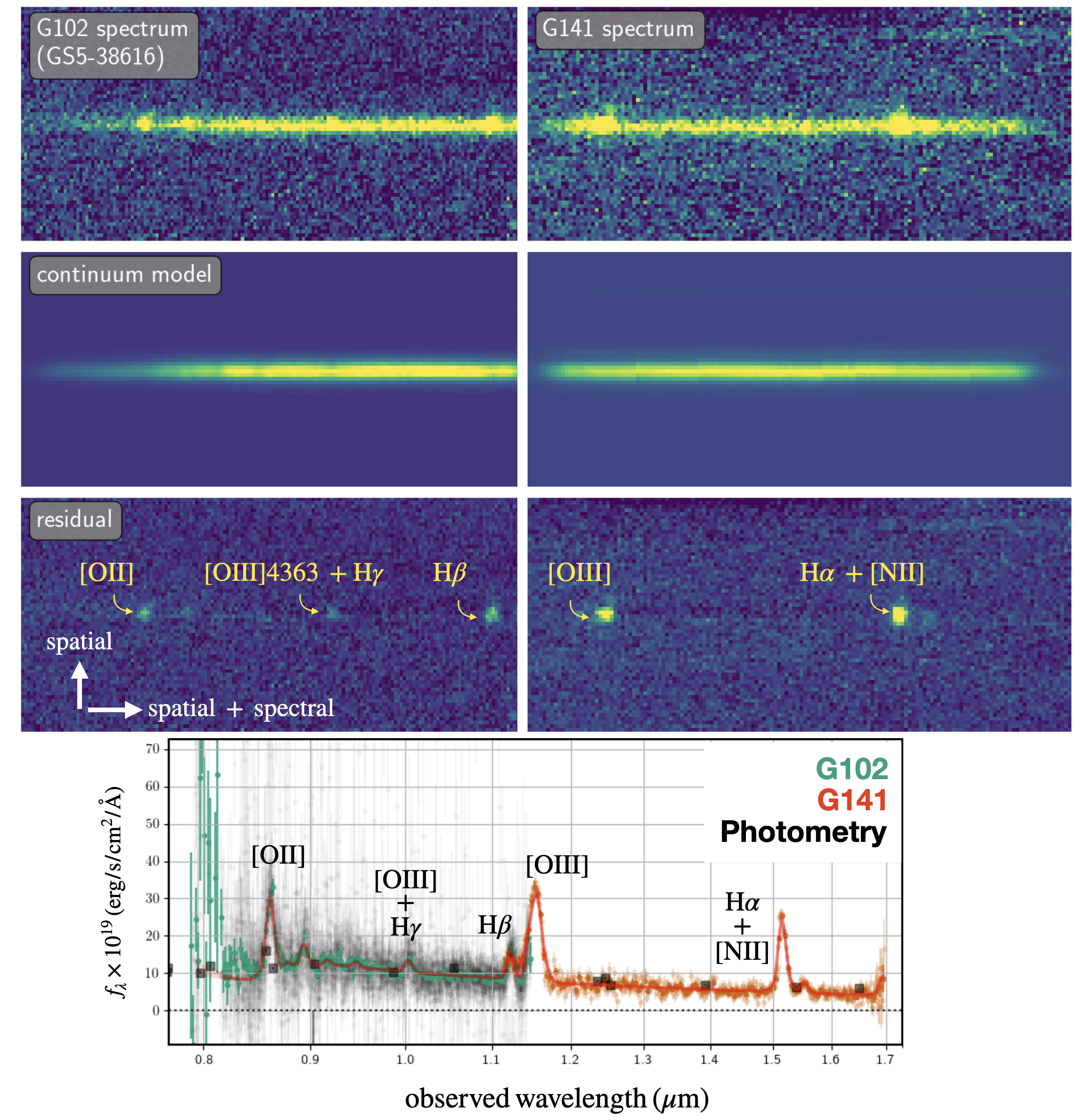}\label{fig:GS5-38616}
\caption{The G102 and G141 spectra of a single galaxy (ID: 38616; z = 1.31) in the GS5 CLEAR field. The top row shows the 2D (spatial $\times$ spatial+spectral) spectra from the WFC3/G102 (left) and WFC3/G141 (right) grisms. The second row shows the respective continuum models. The third row shows the residual of the two (spectrum - continuum). Emission line flux (which is not included in the continuum model) appears as a bright feature in the 2D residual spectrum. Prominent emission line/line complexes are indicated. The bottom panel shows the 1D extracted spectrum (G102 in blue and G141 in red-orange) along with the full continuum + emission line fit (red line). The overlapping photometry is shown with the black squares. The faint grey points show the spectral flux density measured for each individual exposure.}
\end{figure*}

\begin{figure*}
\includegraphics[width=\textwidth]{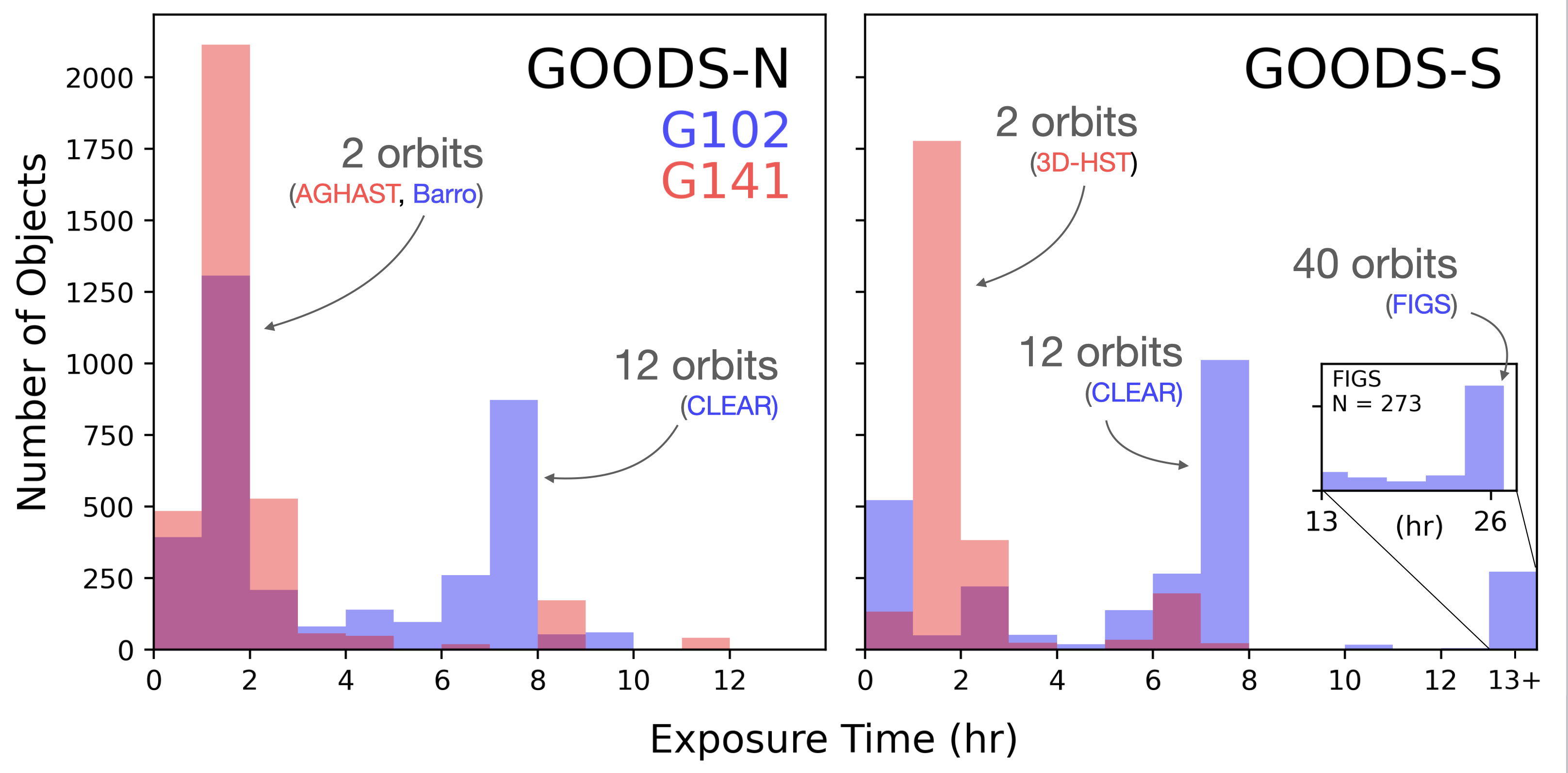}\label{fig:exposure_time}
\caption{The distribution of {\emph{HST}}/WFC3 G102 and G141 exposure times for all objects extracted (m$_{\mathrm{F105W}}\,<\,25$) as a part of the \clearer{} dataset. The peaks in the distribution correspond to observing programs of $\sim$2 orbit (3D-HST, BARRO, AGHAST), 12 orbit (CLEAR), and 40 orbit (FIGS) depth, as indicated.}
\end{figure*}

\subsection{Full-Field Contamination Models}

For each pointing, a contamination model is created to account for spectral overlap of adjacent sources on the WFC3 detector. The contamination model is generated from an iterative forward-model of the full field {\emph{HST}} Y-band mosaic. A first pass model is constructed for all objects in the $Y$-band mosaic brighter than m$_{\text{F105W}}$ $=$ 25. For each object, a spectrum is constructed that is flat in $f_\lambda$ flux density and normalized to the F105W flux of the source. A second pass ``refined'' continuum model is then created for objects brighter than m$_{\text{F105W}}$ $=$ 24. These objects are assigned spectra by fitting 2nd-order polynomials to the spectrum of each source after subtracting the first pass models of suspected contaminating sources. This process is repeated for each visit. The end-to-end reduction and continuum modeling of a single G102 grism exposure is shown in Figure \ref{fig:detector_model}. While the continuum model generally performs remarkably well for the majority of the sources and detector area (see the residual image in the lower right panel of Figure \ref{fig:detector_model}), for point-like sources there can be residual signal due to the imperfect PSF reconstruction in the blotting procedure. To make the grism model, we must blot the more finely-sampled drizzled reference image to the coarser pixels of the detector---where the PSF is undersampled. That transformation is not perfect/lossless, and will not preserve the exact pixel phase sampling. This is most apparent in the residual continuum of bright PSF-sized sources (e.g., stars, AGN). 

We measure the fraction of the extracted source spectra (see next subsections) that are contaminated as a function of the contamination level (F$_{\lambda,\mathrm{contamination}}$/F$_{\lambda,\mathrm{source}}$). We find that $\sim25\,\%$ ( $\sim65\,\%$) of the spectra are contaminated at a level of F$_{\lambda,\mathrm{contamination}}$/F$_{\lambda,\mathrm{source}}$ $\ge$ 1 and 0.1, respectively. In all cases, this continuum contamination is modeled and subtracted.

\subsection{Extraction of Spectra}

We use {\tt{Grizli}} to extract the 2D grism spectra of all objects brighter than m$_{\text{F105W}}$ $=$ 25. Each 2D spectrum is known as a ``beam''. One beam is extracted for each grism observational visit of each object. Therefore, each object normally has multiple beams---one for each PA of each grism instrument. The beam files carry along the local contamination model relevant for the 2D spectrum and a full description of the WFC3 detector. In total, \numextracted{} objects have at least one 2D spectrum extracted from the \clearer{} dataset. Of these, 4707 were observed with both grisms, 533 were observed with only the G102 grism, and 808 were observed with only the G141 grism. 

The grism exposure times of the extracted objects are shown in Figure \ref{fig:exposure_time} and range from $0.5-28$ hours in G102 and $0.5-12$ hours in G141. There are several distinct peaks in the distribution of exposure times, which correspond to different programs in the \clearer{} observational set. The notable peaks indicated in Figure \ref{fig:exposure_time} are associated with programs of depth: $\sim$2 orbit (from Barro/G102, AGHAST/G141, 3D-HST/G141), 12 orbit (CLEAR/G102), and 40 orbit (FIGS/G102), respectively.

\subsection{Redshifts}

Redshift and emission line fits are carried out in {\tt{Grizli}} using both the grism spectra and the available multiwavelength photometry. The spectra are scaled to the photometry using a simple wavelength-independent scaling factor. The continuum is modeled using a basis set of template Flexible Stellar Population Synthesis models ({\tt{FSPS}}; \citealt{2009ApJ...699..486C, 2010ApJ...712..833C}). The {\tt{FSPS}} templates reflect a range of galaxy types and star-formation histories following \citet{2007AJ....133..734B} and \citet{brammer08}. Emission-lines and emission-line complexes are included on top of the {\tt{FSPS}} models.

\begin{figure*}
\includegraphics[width=\textwidth]{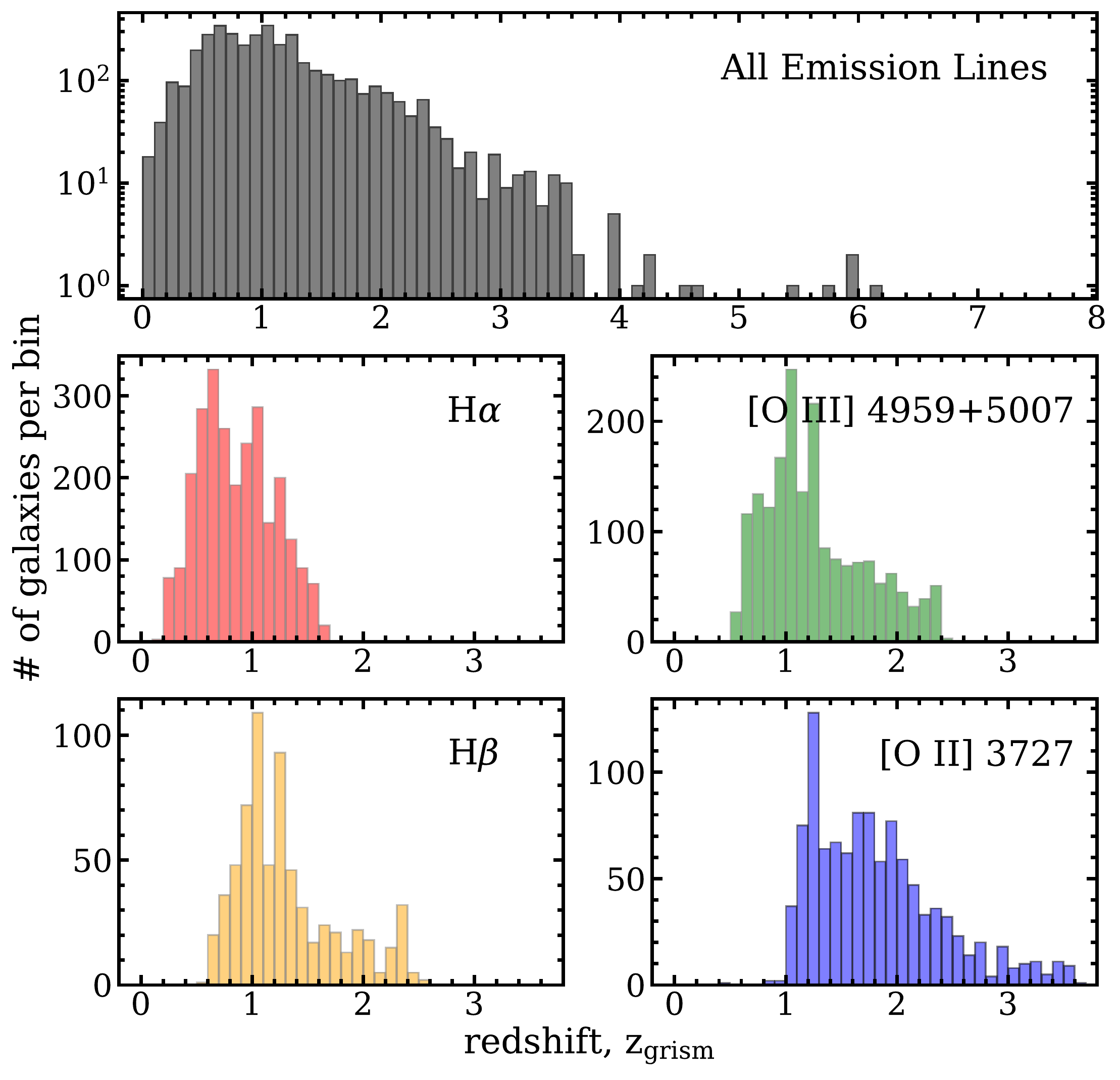}\label{fig:zhist}
\caption{Redshift distribution of galaxies with emission lines measured by CLEAR (in the combined G102 + G141 dataset).  The top panel shows the distribution of all sources with at least one emission line detected with SNR $\ge$ 3 in one of \ha, \hb, \oii, \oiii, \sii, \siii, \mgii, or \lya.  The lower panels show the redshift distribution of sources detected with SNR $\ge$ 3 in a single emission line (as labeled).  The G102 + G141 are sensitive to emission from these lines over different ranges in redshift.}
\end{figure*}

To carry out the redshift fit, the templates are redshifted to a trial redshift and convolved with the bandpass functions of the photometric filters. In this initial redshift fit, the ratios of the emission line complexes are fixed to reduce the redshift degeneracies that would be introduced if the lines were allowed to freely vary. The emission lines/complexes are allowed to freely vary in the final full fit, as described in the next subsection. The redshifted templates are forward-modeled into the observational plane of each extracted 2D spectral ``beam''---using the direct $Y$-band image to define the spatial morphology of the source. This approach accounts for the unique spectral broadening of each galaxy due to its morphology.  The final model is constructed using a non-negative linear combination of the template models. The goodness of fit is computed using the total $\chi^2$ of the 2D spectral pixels and photometry. The uncertainties of the data are taken from the exposure-level noise model and photometric catalog, respectively.  The best redshift is that where the $\chi^2$ is minimized across a grid of redshifts spanning $z\,=\,0$ to $z\,=\,12$. In the top panel of Figure \ref{fig:zhist}, we show the distribution of redshifts of the sample of galaxies with at least one secure line detected (S/N $\geq$ 3). In the bottom panels, we show the distribution for galaxies with line detections in H$\alpha$, H$\beta$, \oiii, and \oii. The majority ($>$95\%) of the galaxies in \clearer{} with redshifts that are based on line detections span the redshift range $0.2 \lesssim z \lesssim 3$.

\subsection{Emission Line Fluxes and Maps}

Emission line fluxes are measured at the best-fit redshift using the basis {\tt{FSPS}} templates including emission lines, and following the forward modeling technique described above. However, now the emission lines and complexes are considered as separate components without fixing their line ratios. The [\ion{S}{2}]$\lambda\lambda$6718+6732 and [\ion{O}{3}]$\lambda\lambda$4960+5007 doublets are fit as single components with line ratios that are fixed at 1:1 and 1:2.98, respectively (\citealt{2006agna.book.....O}). The [\ion{S}{2}] ratio is appropriate for ISM electron densities of $\sim\,10^2-10^3$ cm$^{-3}$ \citep{2019ApJ...880...16K}. The H$\alpha$+[\ion{N}{2}] complex is blended at the resolution of the G141 and G102 grisms.  We therefore fit these lines with a single component at the wavelength of H$\alpha$. The 2D + 1D spectra (G102 and G141) of a single galaxy in the \clearer{} dataset is shown in Figure \ref{fig:GS5-38616}, along with its full {\tt{FSPS}} + emission lines fit. 

Emission-line maps are created by drizzling the continuum- and contamination-subtracted 2D spectral beams to the wavelength of the redshifted line center. This is carried out using the astrometry of the spectral trace. The line maps have a pixel scale of 0\farcs1. The uncertainties on the line maps are computed using the weights of the constituent pixels in the drizzling procedure. Emission line maps are generated automatically for H$\alpha$ + [NII], [OIII]$\lambda$4960,5008, H$\beta$, and [OII]$\lambda$3727,3730. They are created for the remaining lines and line complexes listed in Figure \ref{fig:lines_redshift} if they are detected with a signal-to-noise greater than four in the 1D spectrum. Example line maps created from the CLEAR dataset can be found in \citet{simons21}, \citet{matharu22}, and \citet{backhaus22a, backhaus22b}.

\begin{figure*}
\includegraphics[width=\textwidth]{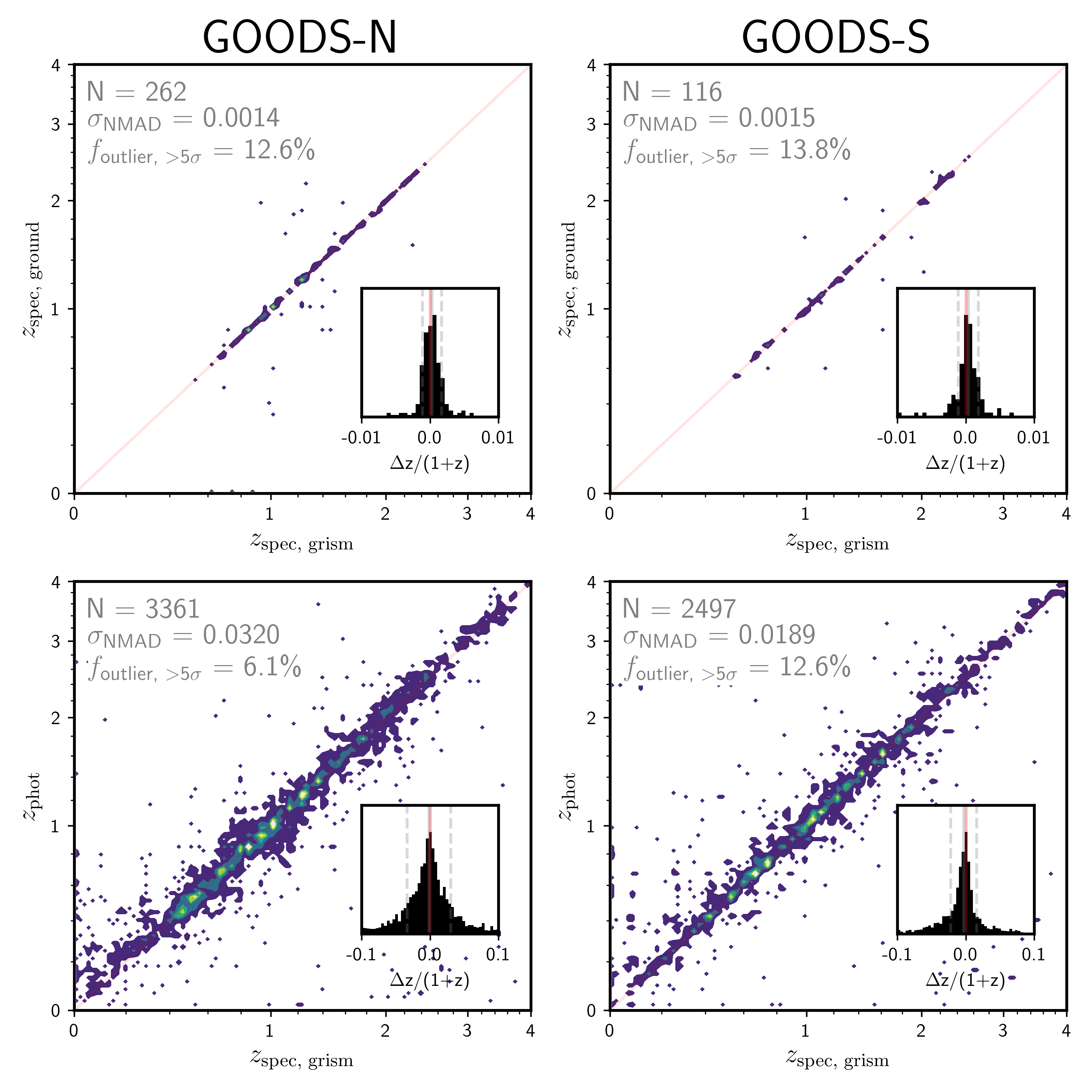}\label{fig:z_versus_z}
\caption{The redshifts measured from the CLEAR grism spectroscopy ($z_{\text{spec, grism}}$, top panels) are compared with those measured from ground-based spectroscopy with high spectral resolution ($z_{\text{spec, ground}}$) where available and those inferred from the photometry alone ($z_{\text{phot}}$, bottom panels). The distribution of redshift differences between the measures are shown in the subpanels. The $z_{\text{spec, grism}}$-$z_{\text{spec, ground}}$ comparison only includes objects with at least one secure line detection in both the grism and ground-based datasets. The redshift uncertainty of the grism spectra is inferred from the width of the distribution of differences with the ground-based redshifts and is quantified $\sigma_{\mathrm{NMAD}}(\Delta\,\mathrm{z}/(1+\mathrm{z}))\sim0.0014$. The FWHM uncertainty is roughly equal to the spectral size of one WFC3/G102 and G141 pixels.  The normalized median absolute deviation (\nmad{}) is shown as a dashed vertical line and is listed in the top left of each panel. The outlier fraction---defined as a redshift discrepancy larger than $5\,\times\,\sigma_{\mathrm{NMAD}}$---is listed in the top left of each panel.}
\end{figure*}

\begin{figure*}
\includegraphics[width=\textwidth]{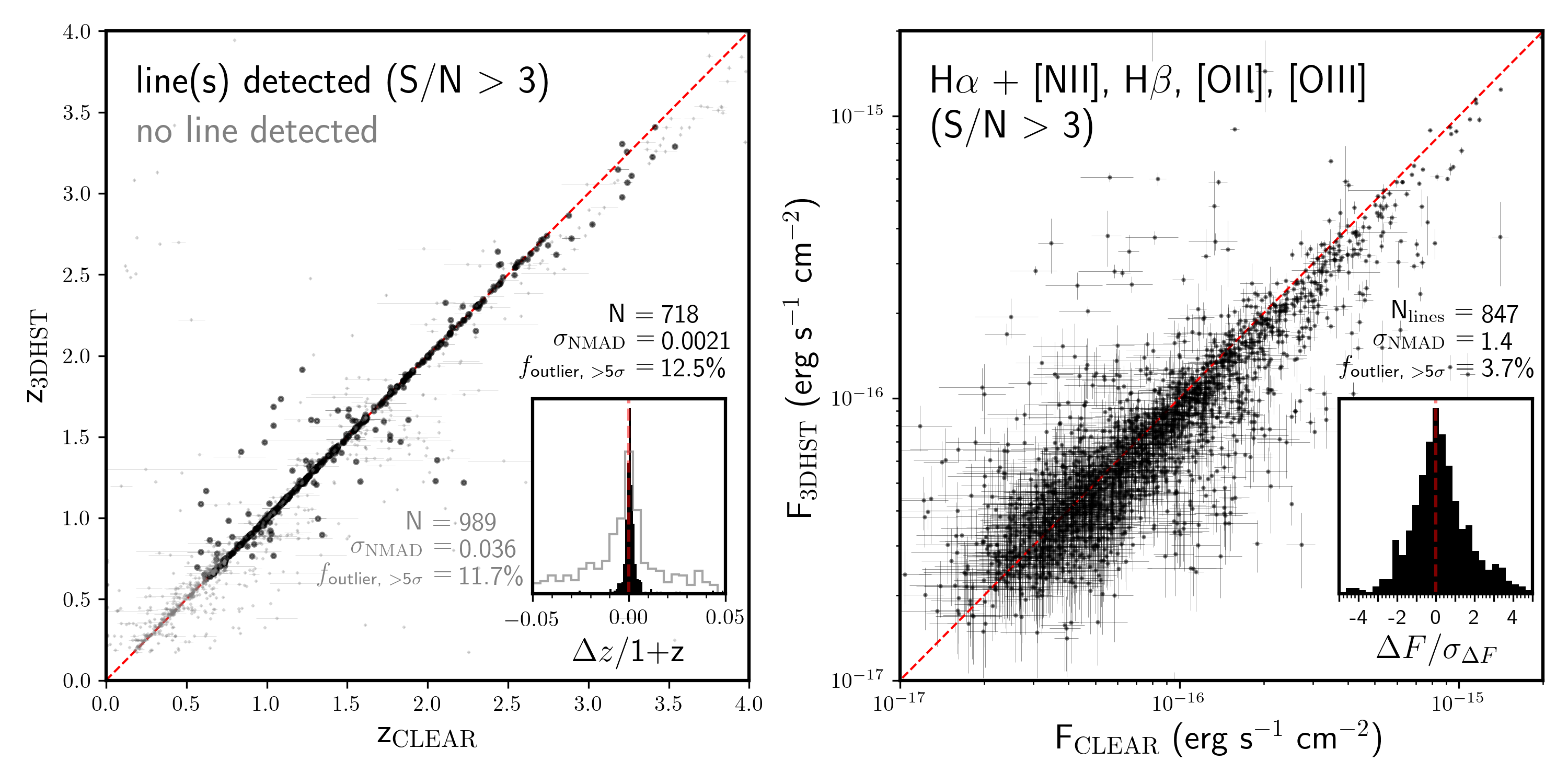}\label{fig:clear_versus_3dhst}
\caption{A comparison of the measured redshifts (left panel) and fluxes (right panel) from the CLEAR and 3D-HST surveys. The G141 observations are identical in both datasets, but the data are reduced using independent codes. CLEAR also contains the G102 data.  The left panel compares the redshifts for sources with at least one line detected (S/N$\,>\,$3) in both reductions (black) and for sources with no line detected (gray). The distribution of differences is shown in the subpanel, with the $\sigma_{\text{NMAD}}$ and outlier fraction listed. For galaxies with a secure line detection, the width of the distribution of differences is very narrow, indicating a level of precision roughly equal to the spectral size of one WFC3 pixel. For those galaxies without a secure line detection, the redshift is effectively the photometric redshift. In these cases the width of the distribution of differences is similar to the bottom panels of Figure \ref{fig:z_versus_z}. The right panel compares the line fluxes for emission lines that are detected with a S/N of 3 or higher in both CLEAR and 3DHST. The distribution of the differences normalized by the uncertainties is shown in the subpanel. }
\end{figure*}

\section{Data Products and Catalogs}\label{section:catalogs} 

This section provides a description and validation of the science products and redshift + line flux catalogs that are produced from the CLEAR survey. The products described here are released alongside this paper at \url{https://archive.stsci.edu/hlsp/clear/}.  An interactive map and the ``biographical'' information for each galaxy in our sample is available at \url{https://clear.physics.tamu.edu}.

\subsection{Data Products} 

As described in \S\ref{section:reduction}, we use the {\tt{Grizli}} grism analysis software to extract spectra and emission line maps for \numextracted{} sources. Each source is associated with a set of four {\tt{Grizli}} products, following a naming scheme [{\tt{FIELD}}]\_[{\tt{ID}}]\_[{\tt{PRODUCT}}].fits. [{\tt{FIELD}}] is the CLEAR field name (e.g., `GN1'; see Table \ref{table:hstobs}), [{\tt{ID}}] is the identification number from the 3D-HST catalog \citep{skelton14}, and [{\tt{PRODUCT}}] is the product type. The product types are `full', `beams', `stack', and `1D'.  These are multi-extension fits (MEF) files, and are described here: 

\begin{itemize}
\item The {{\bf{*\_full.fits}}} product stores the (i) results of the {\tt{FSPS}} + emission line model including the redshift likelihood information (\texttt{ZFIT\_STACK}),  best-fit line fluxes and equivalent widths and covariance elements (\texttt{COVAR}), (ii) the model templates (\texttt{TEMPL}), (iii.) the source segmentation map (extension \texttt{SED}), (iv) the F105W or F141W direct image (extension \texttt{DSCI}) and associated weight map (extension \texttt{DWHT}), and (iv) a map of the point spread function in each WFC3 direct imaging bandpass (extension \texttt{DPSF}). In addition, emission line maps are included for those galaxies studied in \citet{simons21}, \citet{matharu22}, and \citet{backhaus22b}. As described above, the emission line maps are generated for the following line/line complexes if they are in the observed wavelength window of the object: H$\alpha$ + [NII], [OIII]$\lambda$4960,5008, H$\beta$, and [OII]$\lambda$3727,3730. Maps are produced for the remaining lines/line complexes listed in Figure \ref{fig:lines_redshift} only if they are detected in the 1D spectrum with a signal-to-noise greater than four. The maps are 160 pixels $\times$ 160 pixels, which corresponds to 16$\arcsec$ $\times$ 16$\arcsec$ at our pixel scale of 0$\arcsec$.1 $\times$ 0$\arcsec$.1.   For each emission-line that is fit, the MEF contains extensions for the emission-line map (\texttt{LINE}), an associated weight map (\texttt{WHT}), continuum map (\texttt{CONTINUUM}), and contamination map (\texttt{CONTAM}).

\item The {{\bf{*\_beams.fits}}} product stores the full set of G102 and/or G141 grism 2D spectra along with postage stamps of the associated direct reference images. For a G102 spectrum, the corresponding direct image is from the WFC3/F105W filter. For a G141 spectrum, the direct image is from the WFC3/F140W filter. As defined above, an individual 2D spectrum is referred to here as a ``beam". This product serves as the main input to {\tt{Grizli}}'s spectral fitting and emission line map-making tools.  The MEF extensions in this product have the same definitions as for those in the \textbf{*\_stack.fits} products, but the \textbf{*\_beams.fits} contain information for each individual ``beam''.  

\item The {{\bf{*\_stack.fits}}} product stores a stacked 2D spectrum of the beams, including the science extension (\texttt{SCI}), a weight extension (\texttt{WHT}), a contamination model extension (\texttt{CONTAM}), a best-fit continuum model extension (\texttt{MODEL}), and an estimate of the point-spread function (\texttt{KERNEL}).  

\item The {{\bf{*\_1D.fits}}} product stores the optimally-extracted 1D grism spectrum of the source. There is one MEF extension for each of the G102 and G141 spectra.  Each of the fits extensions of this product includes columns of the wavelength (``wave''), (unnormalized) flux density (``flux''), flux--density error (``err''), number of grism spectral pixels per wavelength bin (``npix''), flat (``flat''; used for normalization of the flux), contamination model (``contam''), and a decomposition of the spectrum into its line (``line'') and continuum (``cont'') components.  To convert the unnormalized spectrum to flux--density (in units of erg s$^{-1}$ cm$^{-2}$ \AA$^{-1}$) one divides the `` flux'' column  by the ``flat'' column.  

\end{itemize}

\subsection{Line Fluxes and Redshifts} 

Here we describe the line flux and redshift catalogs that are released alongside this paper. We also carry out a relative validation of the redshifts and line fluxes by comparing them against a compilation of high spectral resolution redshifts from ground-based spectroscopic surveys and previous grism-based measurements from the 3D-HST team \citep{momcheva16}.

\subsubsection{Catalogs}
The redshifts and line fluxes that are measured from the \clearer{} dataset are released in two spectroscopic catalogs: one for GOODS-S (``GDS\_v4.1\_CLEAR.fits') and one for GOODS-N (``GDS\_v4.1\_CLEAR.fits''). The catalog version released alongside this paper is \linecatversion{}. The columns of these catalogs are listed in Table~\ref{table:grizli_catalog}. 

The catalogs include basic properties of the galaxies and the grism observations: the source ID (identical to those of 3D-HST; \citealt{skelton14}), the J2000 ICRS right ascension and declination, the number of emission lines/line complexes observed by the grisms,  the on-source G102 and G141 exposure time, and diagnostics of the template fit including the minimum $\chi^2$ and the Bayesian information criterion BIC\_TEMP. 

The catalogs also include the redshift and emission line measurements from {\tt{Grizli}}: the confidence intervals of the redshift probability distribution, the maximum likelihood\footnote{{\tt{Grizli}} adopts the maximum likelihood redshift to create emission line maps from the grism spectra.} and minimum ``risk'' redshifts \citep{2018PASJ...70S...9T}, the line flux and line flux uncertainties for the full suite of lines listed in Figure~\ref{fig:lines_redshift}, and the confidence intervals of the rest-frame equivalent widths of these lines.

\subsubsection{Comparison and Validation of Redshifts and Line Fluxes}

In Figure \ref{fig:z_versus_z}, we compare the redshifts measured from the \clearer{} dataset against those obtained for the same galaxies from: (1) ground-based spectra with a factor of $\sim\times10$ higher spectral resolution than the HST grisms and (2) fits to the photometry alone (as described in \S\ref{section:photometry}). The former are known as ``spectroscopic redshifts" in the 3D-HST catalogs and the latter are known as ``photometric redshifts". 

The \clearer{} redshifts blend these approaches. They are measured combining the information from {\emph{both}} the low-spectral resolution grism spectra---which carry diagnostic emission line and continuum information---and the broadband photometry. For galaxies with little valuable information provided by the grism spectra (i.e., no emission lines or continuum breaks detected), the redshifts are effectively derived from photometry alone. In those cases, the accuracy of the redshifts will be similar to those of the ``photometric redshifts''. On the other hand, for galaxies with emission lines detected in the grism spectra, the uncertainty of the derived redshifts is generally much smaller. With an emission line detection, the redshift precision will only be limited by the ability of the grism data to centroid the emission feature(s)---which is set by the spectral resolution and pixel sampling.

\vspace{0.2cm}
\noindent{\bf{\underline{\clearer{} vs. ground-based redshifts}:}} In the top panel of Figure \ref{fig:z_versus_z}, we compare the ground-based spectroscopic redshifts against those measurements from the joint \clearer{} grism + photometry dataset. The ground-based redshifts ($z_{\mathrm{spec, ground}}$) are sourced from three matched catalogs: the compilation provided in the original 3D-HST catalog in GOODS-N and GOODS-S (\citealt{skelton14}; N = 1206), the KMOS-3D survey in GOODS-N (\citealt{2019ApJ...886..124W}; N = 43), and the MOSDEF survey in GOODS-N and GOODS-S (\citealt{2015ApJS..218...15K}; N = 143). For the KMOS-3D and MOSDEF surveys, we include redshifts with quality flags of $>=1$ (described as a secure redshift) and $>=3$ (described as a redshift based emission line(s) detected at S/N of 2 or better), respectively.

We only consider sources with at least one secure line detected in the \clearer{} dataset (S/N $>$ 3). In both fields, we find excellent systematic agreement between the two redshifts, with a distribution of differences that is centered on $\Delta\,\mathrm{z}/(1+\mathrm{z})\,\sim\,0$ (shown in the subpanel). The median of $\Delta\,\mathrm{z}/(1+\mathrm{z})$ is 0.0002 $\pm$ 0.0001 and 0.0003 $\pm$ 0.0002 in GOODS--N and GOODS-S, respectively---i.e., the median is statistically consistent with 0.

Given the high spectral resolution of the ground-based data ($\times10$ higher than the WFC3 grisms), the width of this distribution is effectively an exclusive measure of the redshift precision of the grism data. To quantify the level of that precision, we measure the normalized median absolute deviation (\nmad{}) of the redshift differences (following \citealt{brammer08, momcheva16}):

\begin{equation}
\sigma_{\mathrm{NMAD}}\,=\,1.48\times\mathrm{median}\,\left(\left|x - \mathrm{median}\,(x)\right|\right)
\end{equation}

\noindent where $x = \Delta z/(1+z_{\mathrm{spec,grism}})$. The quantity \nmad{} is the median absolute deviation multiplied by a factor of 1.48. For a normal distribution, \nmad{} is equal to the standard deviation. However, it is less impacted by outliers than the standard deviation. 

The \nmad{} of the differences between the ground-based redshifts $z_{\mathrm{spec, ground}}$ and the grism redshifts $z_{\mathrm{spec, grism}}$ is $\sim$0.0014 for both fields. The implied full width at half-maximum of the redshift accuracy is $0.0033\,\times\,(1+z)$. This corresponds to 33 and 46 \AA\, at 1.02 $\mu$m and 1.41 $\mu$m, respectively---the characteristic wavelengths of the WFC3 grisms. These are roughly equal to the spectral size of one WFC3 pixel for both the G102 grism (24.5 \AA) and the G141 grism (46 \AA). To conclude, the FWHM redshift accuracy of the WFC3 grisms is roughly equal to their spectral sampling size. This is generally consistent with that found for the G141 grism in the 3D-HST survey \citep{momcheva16}. 

Next, we measure the outlier fraction of the distribution of differences. We define an outlier as a galaxy with a difference in the redshift measurements that is larger than $5\,\times\,\sigma_{\mathrm{NMAD}}$. We measure an outlier fraction of $\sim$13$\%$ in both fields. Roughly half of the outliers have a discrepancy that is statistically consistent (i.e., within $\,5\,\times\,\sigma_{\mathrm{NMAD}}$) with that expected from simple line confusion between the surveys---e.g., a redshift based on a line identified as H$\alpha$ in one dataset and \oiii{} in another. For the remaining outliers (24 of 378 total, or 6$\%$ of the full sample), the $>5\,\sigma$ discrepancies are unexplained. A potential explanation is that the grism spectra of these galaxies could be contaminated by one or more emission lines from another source---which could lead to a spurious spectroscopic redshift based on those imposter lines. Such line contamination is not accounted for in the {\tt{Grizli}} models. Of the 24 unexplained outliers, 13 have redshifts constrained by multiple secure line detections ($5\sigma$) in the grism spectra. These objects (13 of 378, or 3$\%$ of the full sample) can not be easily explained by grism line contamination.

\vspace{0.2cm}
\noindent{\bf{\underline{\clearer{} vs. photometric redshifts}:}} In the bottom panel of Figure \ref{fig:z_versus_z}, we compare the \clearer{} redshifts with those estimated using only the photometry ($z_{\mathrm{phot}}$). As above, we find excellent systematic agreement between the two---in both fields the medians of the distributions of redshift differences are statistically consistent with 0. The \nmad{} of the differences is 0.0320 and 0.0189 for GOODS-N and GOODS-S, respectively. Those are a factor of $\sim\times\,10-20$ larger than that measured above between the grism and ground-based spectroscopic redshifts. We conclude that the redshifts based on photometry-alone are $\sim10-20\,\times$ less precise than those based on the grism + photometry.

\vspace{0.2cm}
\noindent{\bf{\underline{\clearer{} vs. 3D-HST redshifts and line fluxes}:}} In Figure \ref{fig:clear_versus_3dhst}, we compare the \clearer{} redshifts and emission line fluxes with those measured by the 3D-HST survey for the same galaxies \citep{skelton14,brammer14,momcheva16}. The 3D-HST G141 observations are also included in the \clearer{} dataset, and so this comparison is effectively a test of the differences between the 3D-HST and {\tt{Grizli}} reduction pipelines.

In the left panel of Figure \ref{fig:clear_versus_3dhst}, we compare redshifts for two types of sources: (1) those with a secure line detected in both surveys (black), and (2) those with no line detected in either survey (gray). The distribution of differences is shown in the subpanel. The systematic agreement between the two survey measurements is excellent, peaking at $\Delta\,z/(1+\mathrm{z})\,\sim\,0$ (with a median that is statistically consistent with 0). As expected, we find that the distribution of differences for those galaxies with a secure line detection is much more narrow (\nmad{}$\sim$0.002) than for those without one (\nmad{}$\sim$0.036). 

In the right panel of Figure \ref{fig:clear_versus_3dhst}, we compare the measured fluxes for the bright Balmer and oxygen rest-optical emission lines that are securely detected (S/N$\,>\,$3) in both surveys. In the subpanel, we show the distribution of the differences normalized by the uncertainty of the differences ($\Delta\,F/\sigma_{\Delta\,F}$). The quantity $\sigma_{\Delta\,F}$ is calculated as ($\sigma_{F,1}^2 +\sigma_{F,2}^2$)$^{1/2}$. If the quoted uncertainties of the individual measurements reflect the true uncertainty and there is no systematic offset between the measures, the quantity $\Delta\,F/\sigma_{\Delta\,F}$ should be distributed as a standard normal ($\sigma_{NMAD}\,\sim\,1$, centered on 0).  For the full sample of line detections shown in Figure \ref{fig:clear_versus_3dhst}, we find that the peak of the distribution of $\Delta\,F/\sigma_{\Delta\,F}$ is $\sim$0, indicating systematic agreement, but that \nmad{}$\sim$1.4. At face value, the fact that \nmad{} is larger than 1 could indicate that the individual uncertainties are generally under-estimated. 

However, we note that at the brighter end of the sample the {\tt{Grizli}}-derived fluxes F$_{\mathrm{CLEAR}}$ from the \clearer{} dataset are generally larger than those measured using the 3D-HST pipeline F$_{\mathrm{3DHST}}$. We explore this further by dividing the full sample into bright and faint lines, defined as F$_{\mathrm{CLEAR}}\,<\,10^{-16}$ and $>10^{-16}$ erg s$^{-1}$ cm$^{-2}$, respectively. For the brighter lines, we find that the \clearer{} fluxes are systematically higher than the 3D-HST fluxes by $\sim$0.05 dex. For the fainter lines, we find general systematic agreement with no offset on average. Splitting by line flux, \nmad{} of the fainter lines ($<\,10^{-16}$ erg s$^{-1}$ cm$^{-2}$) is 0.96 while for the brighter lines ($>\,10^{-16}$ erg s$^{-1}$ cm$^{-2}$) it is 1.86. This indicates that the larger \nmad{} for the full population is fully driven by the discrepancy at the brighter end.

In summary, the {\tt{Grizli}}-derived redshifts measured from the \clearer{} dataset are generally consistent with earlier measurements from the ground and from the 3D-HST survey \citep{2016ApJS..225...27M}. We find an overall redshift precision of the grism of \nmad{}$\,=\,0.0014$ in $\Delta z/(1+z)$ for galaxies with a secure emission line detected. For bright emission lines ($>\,10^{-16}$ erg s$^{-1}$ cm$^{-2}$), we find that the {\tt{Grizli}}-derived line fluxes derived as a part of the \clearer{} processing are $\sim0.05$ dex higher than those measured from the 3D-HST pipeline---using the same G141 grism data. However, for faint lines ($<\,10^{-16}$ erg s$^{-1}$ cm$^{-2}$) which comprise the slight majority (70$\%$) of the \clearer{} line detections, the {\tt{Grizli}}-derived line fluxes are in excellent systematic agreement with those from 3D-HST.

\begin{figure*}
\centering
\includegraphics[width=0.9\textwidth]{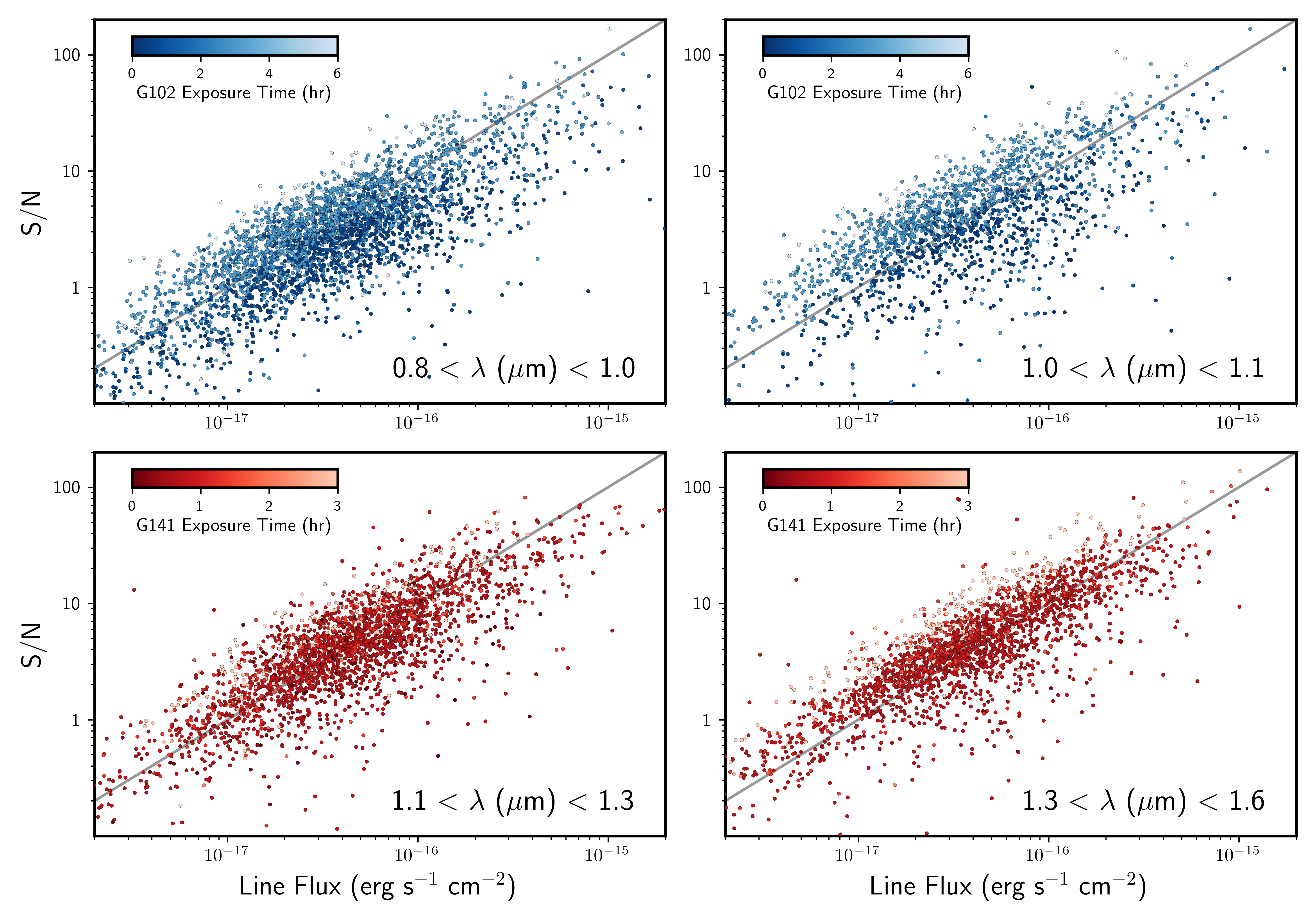}\label{fig:line_flux_snr_exptime}
\caption{The signal-to-noise of the emission lines observed through the \clearer{} observations are shown as a function of line flux, observed wavelength, and grism exposure time. The 5$\sigma$-depth of the emission line fluxes range from $\sim2\,\times\,10^{-17}$ to $1\,\times\,10^{-16}$ erg s cm$^{-1}$ (see Figure \ref{fig:five-sigma_exposure}).}
\end{figure*}

\begin{figure*}
\centering
\includegraphics[width=0.9\textwidth]{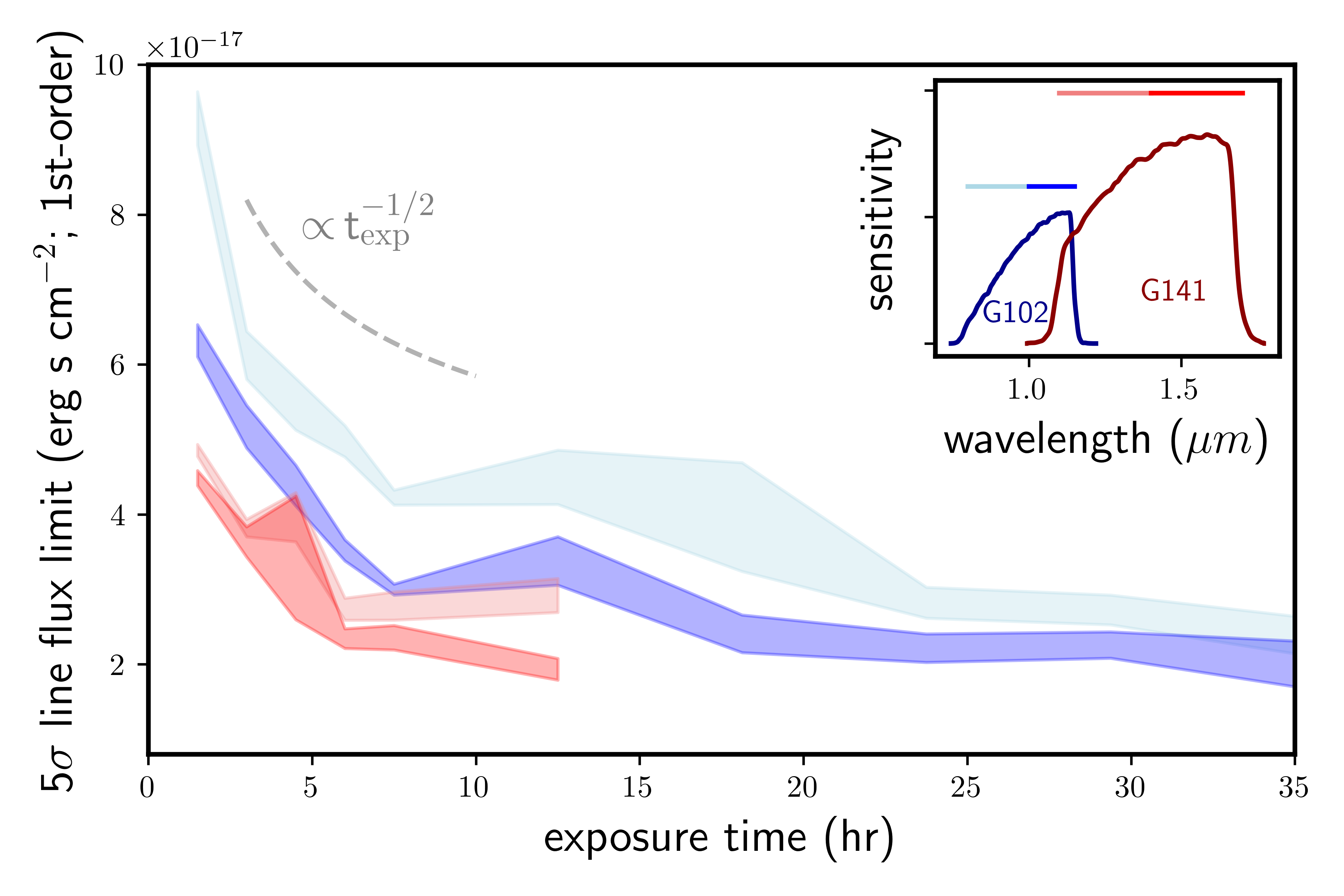}\label{fig:five-sigma_exposure}
\caption{The 5$\sigma$ limit of the emission line flux of the \clearer{} grism spectra is shown as a function of on-source exposure time and the spectral portion of the WFC3 grisms used. The limits are empirically-derived from the full \clearer{} dataset.  These are mostly similar to the theoretical expectation that the flux limit will decrease as the square-root of the exposure time, limit $\propto t_\mathrm{exp}^{-{1/2}}$, as illustrated by the dashed gray line. The relative sensitivity curves of the G102 and G141 grisms are shown in the top right subpanel \citep{kuntschner10}. The different spectral portions of the grisms are indicated with the color-coding defined in the subpanel.}
\end{figure*}

\begin{figure*}
\includegraphics[width=\textwidth]{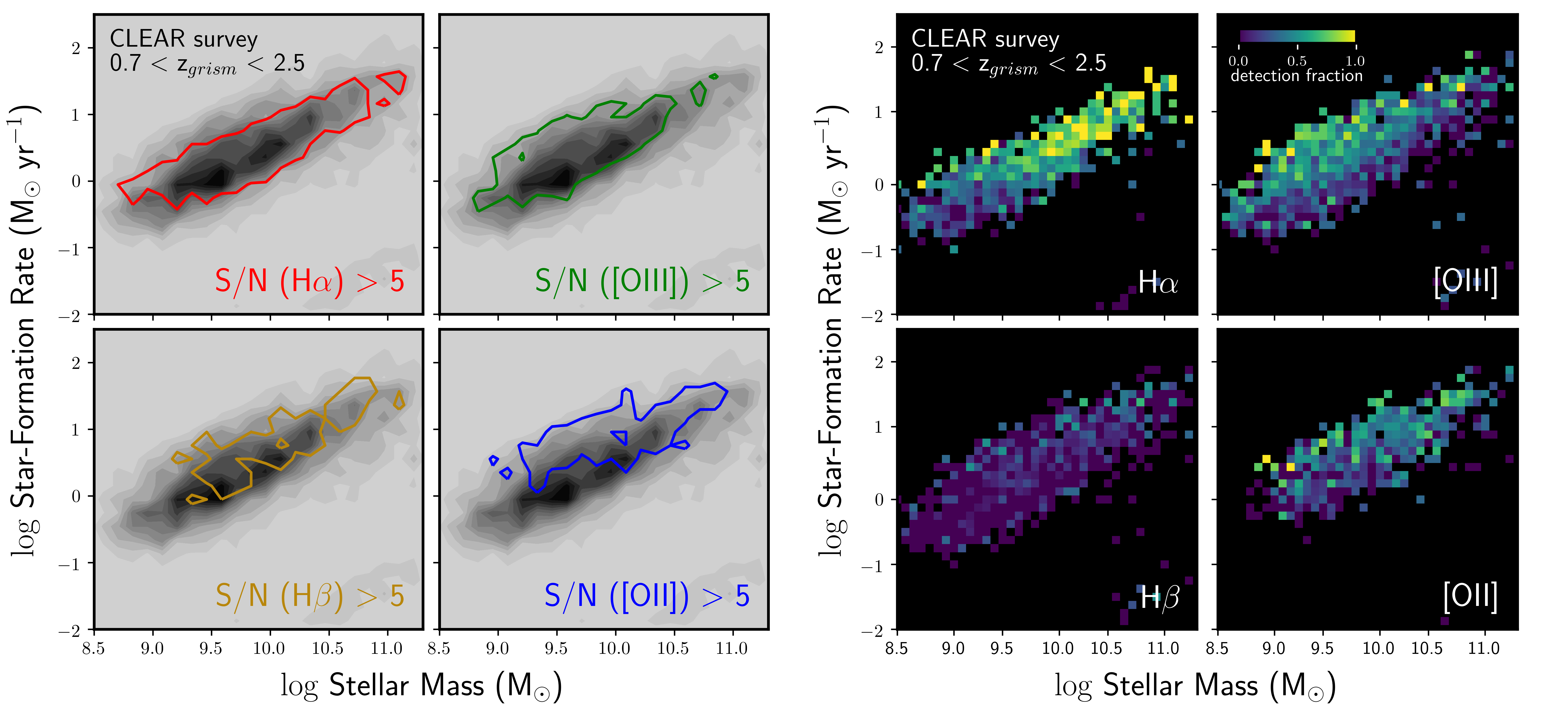}\label{fig:mstar_sfr_combined}
\caption{The emission line detection rate of the CLEAR dataset is shown in the plane of galaxy stellar mass and star-formation rate. Left panel: the full CLEAR galaxy sample (m$_{\mathrm{F105W}}$ < 25) is shown in black and the galaxies with emission line detections (S/N $>$ 5) in each of the indicated lines is shown by the color contours. Right panel: the fraction of galaxies in the full CLEAR sample with an emission line detected (i.e., $\mathrm{N}_{>5\,\sigma}\,/\,\mathrm{N}_{\mathrm{total}}$) in each of the indicated lines. $\mathrm{N}_\mathrm{total}$ is the total number of galaxies in CLEAR with observing conditions that allow for a potential line detection---i.e., for each object, it is considered whether the observed wavelength of the line overlaps with that available G102/G141 grism coverage.}
\end{figure*}

\subsubsection{Line Flux Limits}

In this subsection, we describe the emission line flux depth of the \clearer{} spectra. The line sensitivity of the grisms depend on three factors: (1) the on-source observing time, (2) the spatial extent of the emission, and (3) the observed wavelength of the emission line. 

The sensitivity is lower for galaxies with more extended emission, which distribute over a larger number of pixels on the WFC3 detector and collect more {\emph{spatially-integrated}} noise per observing time\footnote{The emission flux is distributed on the WFC3 detector over $N_{\mathrm{pix}}\,\propto\,R^{2}$ pixels. The spatially-integrated noise scales as $\sigma\,\propto\,N_{\mathrm{pix}}^{1/2}\,\propto\,R$.}. The line sensitivity also depends on the wavelength-dependent throughput of the grisms (see inset of Figure \ref{fig:five-sigma_exposure}). For both the WFC3/G102 and G141 grisms, the throughput is generally higher at the redder end. The G141 grism is roughly twice as sensitive as the G102 primarily because the spectral resolution is twice as low \citep{2011wfc..rept....5K}. With the G141 grism spectra taken through the 3D-HST survey, \citet{momcheva16} show that the line uncertainty scales linearly with the size of the observed galaxy and as the squared-inverse with the throughput. 

As follows, we explore the line flux limits in the \clearer{} dataset. By its construction, this dataset contains programs with a range of observing times (Figure \ref{fig:exposure_time}) and that use one or both of the WFC3 grisms. As such, we want to quantify the line flux limit as a function of on-source exposure time and observed wavelength.

In Figure \ref{fig:line_flux_snr_exptime}, we show the signal-to-noise of the emission lines measured in \clearer{} as a function of line flux, exposure time, and observed wavelength. The gray line in each panel is fixed. 

The wavelength and exposure time dependence is quantified explicitly in Figure \ref{fig:five-sigma_exposure}, which shows the empirically-derived 5$\sigma$ emission line depth for the lines in four grism wavelength windows. At a given exposure time and wavelength, the S/N of the observed lines is higher at the redder end of each of the grisms. For low exposure times ($<4$ hours), the 5$\sigma$ depth ranges from $\sim\,8\times10^{-17}$ erg s cm$^{-2}$ at the blue end of the G102 grism, to $\sim\,4\times10^{-17}$ erg s cm$^{-2}$ in the G141 grism. The latter is consistent with the emission line limit derived from the full sample of 2-orbit G141 data from the 3D-HST survey \citep{momcheva16}. For the deepest G102 data included in this paper (which are that way in large part because they include the ultra-deep observations from the FIGS survey, \citealt{pirzkal17, pirzkal18}), the emission line depth reaches $\sim2\times10^{-17}$ erg s cm$^{-2}$. In general, the redder G141 grism is twice as sensitive as the G102 grism and the redder end of each grism is more sensitive than their bluer end. 

In Figure \ref{fig:mstar_sfr_combined}, we show how the depth of the CLEAR dataset maps onto the plane of star-formation rate versus stellar mass for star-forming galaxies. The \ha\ and \oii\ emission lines are detected in more than 50$\%$ of galaxies with star-formation rates $>$ 1 M$_{\odot}$ yr$^{-1}$ over the redshift range $0.7\,<\,z\,<\,2.5$. The weaker H$\beta$ line is generally detected only in massive and more star-forming galaxies. The doubly-ionized [OIII] line is preferentially detected in the upper end of the main sequence. The [OIII] line is expected to be brighter above the main sequence than below it, given the scaling between the specific star-formation rate and the ionization parameter \citep{papovich22}.

\begin{deluxetable*}{rcl}
\tablecolumns{3}
\tablewidth{0pt}
\tablecaption{Description of the Spectroscopic Catalog\label{table:grizli_catalog}}
\tablehead{
\colhead{column} & 
\colhead{units} & 
\colhead{description}}
\startdata
  ID & & Galaxy ID, matched to \citet{skelton14} \\
  RA & [deg]  & right ascension, J2000  \\
  DEC &[deg]  & declination, J2000  \\
  nlines & &number of emission lines observed with grism\\
  z$\_{(02, 16, 50, 84, 97)}$ & & Confidence intervals of redshift\\
   z$_{{\mathrm{MAP}}}$ & & maximum likelihood redshift\\
   z$_{{\mathrm{RISK}}}$ &  & redshift at minimum ``risk" (defined in \citealt{2018PASJ...70S...9T})\\
   {\tt{[LINE]}}\_FLUX & [1e-17 erg/s/cm$^2$] &line flux \\
   {\tt{[LINE]}}\_FLUX\_ERR & [1e-17 erg/s/cm$^2$] & uncertainty of line flux\\
   {\tt{[LINE]}}\_EW\_RF\_(16/50/84) & [\AA] & percentiles of rest-frame equivalent width\\
   (T$_{\mathrm{G102}}$, T$_{\mathrm{G141}}$) & [s] & observing time\\
   BIC\_TEMP & &Bayesian information criterion of template fit\\
   CHIMIN & &minimum of $\chi^2$ function\\
   DOF & & degrees of freedom\\
  \enddata
        \tablenotetext{a}{{\tt{[LINE]}} is the name of the emission line.}
\end{deluxetable*}

\section{Science Goals and Results}\label{section:science}

The original CLEAR proposal listed several primary science goals, including using the grism dataset: (1.)\ to measure spectroscopic redshifts for hundreds of galaxies in the redshift range $1 < z < 8$ to fainter limits than possible from ground-based spectroscopy;  (2.)\ to provide a measurement of the \lya\ distribution function; and (3.)\ to measure the evolution in the \lya\ equivalent-width distribution function as a test of models of reionization.    CLEAR also contained many secondary science goals as the dataset provides spectroscopic coverage of features from stellar populations and nebular regions from the rest-frame UV to near-IR (depending on the galaxy redshift).   Those secondary goals included studying: (4.)\ the properties of stellar populations in galaxies at $1 \lesssim z \lesssim 2$, (5.)\ the gas-phase metallicities and gas ionization in galaxies at $z > 1$, (6.)\ star-formation in galaxies via the Hydrogen recombination lines, e.g., \ha, \hb, \pab, and (7.)\ spatially-resolved emission in galaxies at $z > 0.5$. Largely these goals have been realized. In what follows, we summarize the key science results enabled by the CLEAR team to date, and how these relate to the original science goals. 

\subsection{Measurement of Hundreds of Galaxy Redshifts}

The \hst/WFC3 G102 and G141 spectroscopy cover emission line and absorption features in galaxy spectra that enable accurate spectroscopic redshifts for galaxies in the CLEAR fields. In total, the \clearer{} dataset constrains the spectroscopic redshifts of 3900 galaxies in GOODS-N and GOODS-S using the detection(s) (SNR $>$ 3) of the integrated emission of one or more emission lines (including \lya, \mgii, \oii, \hb, \oiii, \ha+\nii, \sii, \siii, or \pab).   Figure~\ref{fig:zhist} shows the distribution of galaxies with redshifts measured from at least one emission line.  The redshift distribution has a median of $\langle z \rangle = 1.02$ with an interquartile range (25-75th percentile) of 0.68 to 1.44.  There is a long tail to higher redshift, where the redshift of the 95th percentile extends to 2.42.

Figure~\ref{fig:zhist} also shows the distribution of galaxies with detections of \ha+\nii, \oiii, \hb, or \oii.  Only galaxies with SNR $>$3 in the labeled emission line are included. The redshift span of these subsamples shifts with the rest wavelength of the respective lines (as the different lines are detected in the grism spectra over different redshift ranges, see Figure~\ref{fig:lines_redshift}), with \ha+\nii\ being available for $z\sim 0.5-1.5$, \oiii\ for $z\sim 0.7-2.4$, \hb\ for $z\sim 0.6 - 2.3$, and \oii\ for $z\sim 1.1$ to $z > 3$ in some cases. We detect (at SNR $>$3) 1724 galaxies in \ha+\nii, 1225 galaxies in \oiii, 678 galaxies in \hb, and 1076 galaxies in \oii. Note that these are not \textit{unique} samples as some galaxies are detected in multiple lines at $>3\sigma$ significance.

\begin{figure*}
\centering
\includegraphics[width=\textwidth]{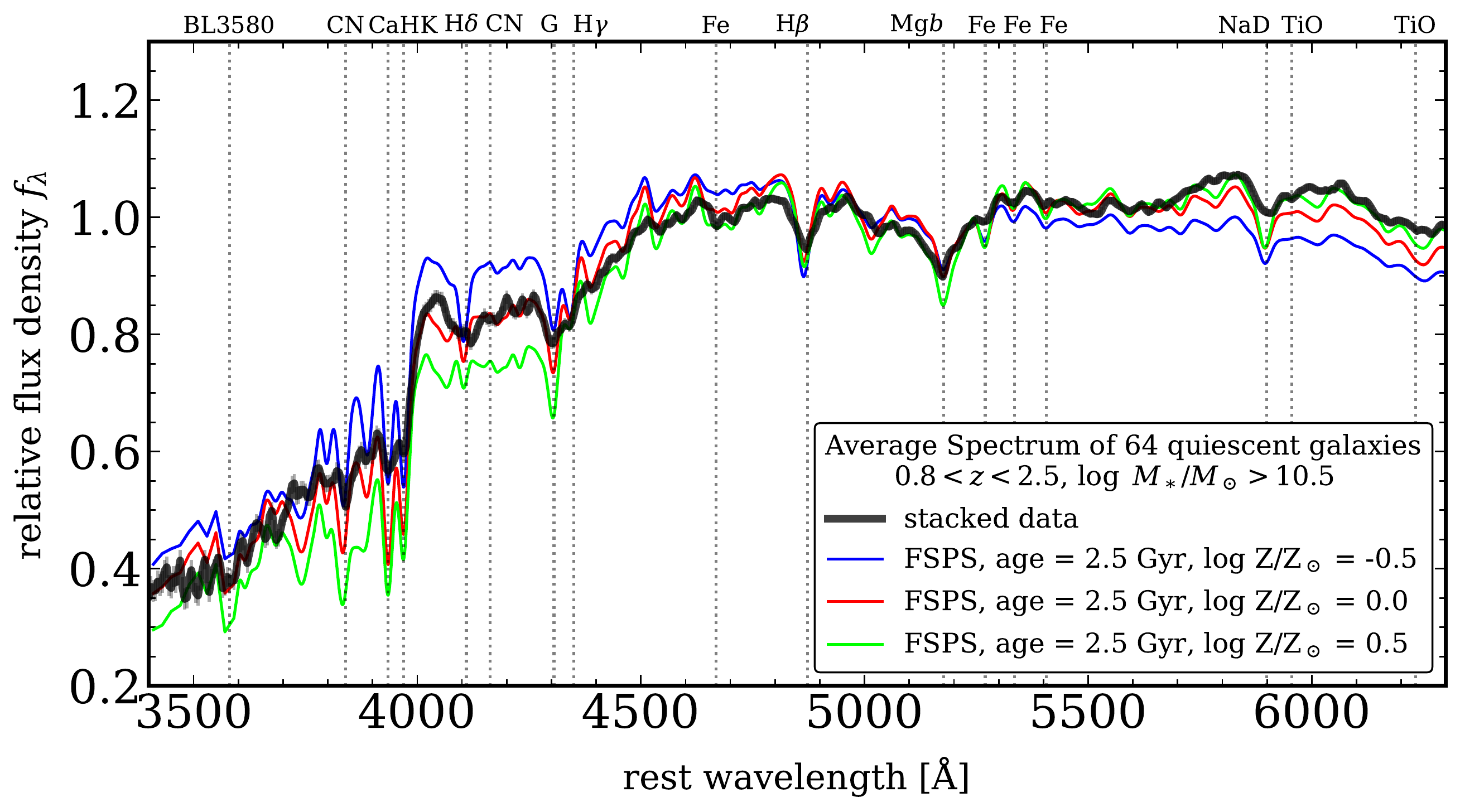}
\caption{The rest-frame average spectrum of quiescent galaxies at $0.8 < z < 2.5$ with stellar mass $\log M_\ast/M_\odot > 10.5$ in the CLEAR fields is shown.  The spectrum shows the G102 and G141 data for 64 galaxies selected to be quiescent (based on the $UVJ$ rest-frame colors) and shifted to the rest-frame wavelength using their measured grism redshifts (heavy black line).  The spectrum of each galaxy is normalized in the rest-frame wavelength range, 4600--5500~\AA, before taking the median flux density for the stack.  The error bars show the standard deviation in the sample.  The dotted lines show wavelengths of prominent absorption lines and spectral indices (see \citealt{worthey1994}).  The thin, colored lines show simple stellar population models (FSPS, \citealt{conroy2009,conroy2010}), formed in an instantaneous burst with an age of 2.5 Gyr, and metallicity (as indicated, see legend), binned to $R\sim 200$.  } \label{fig:quiescent_example}
\end{figure*}

\begin{figure*}
\includegraphics[width=\textwidth]{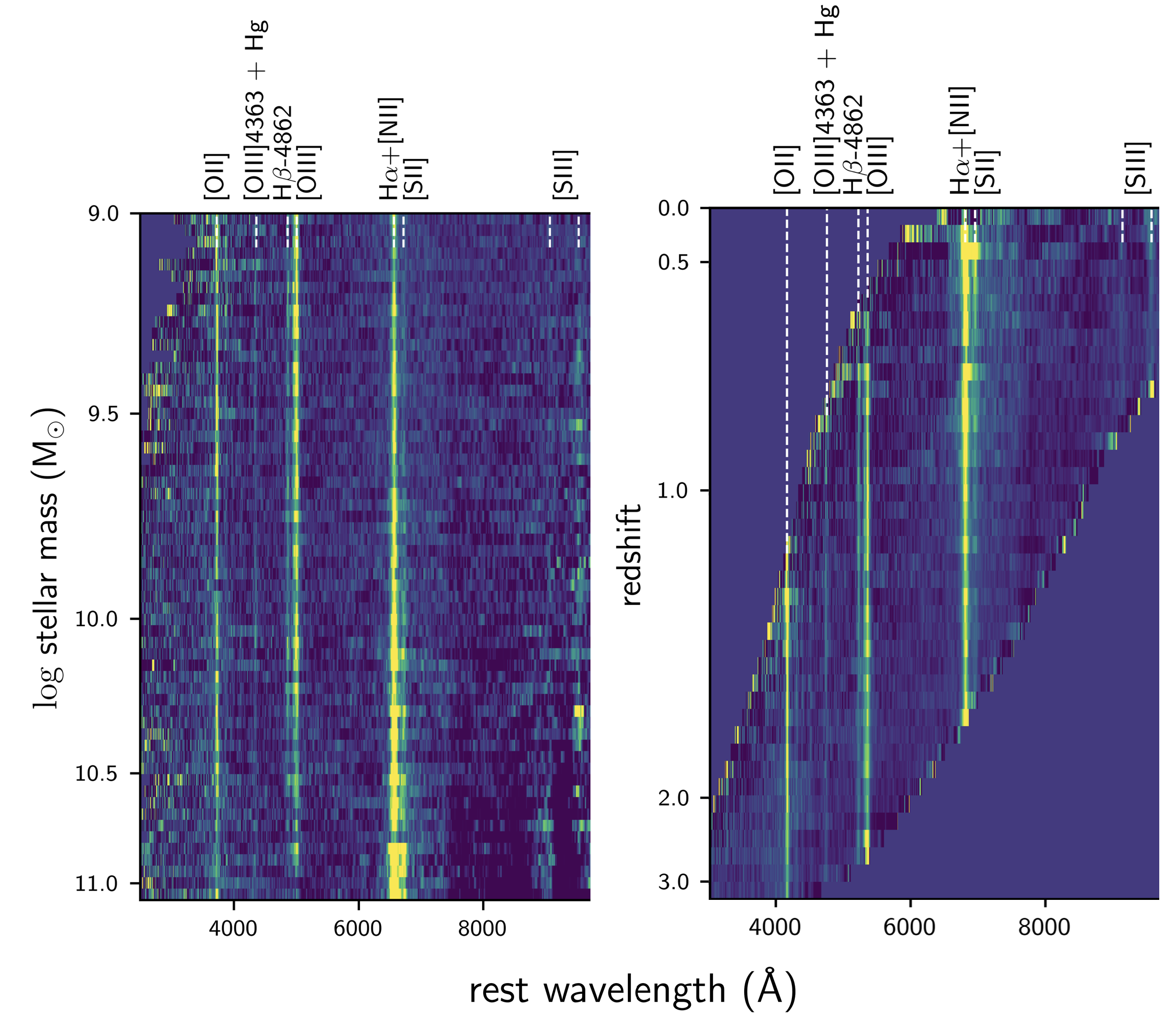}\label{fig:stack_heatmap}
\caption{Stacks of the CLEAR 1D spectra are shown in bins of stellar mass (left panel) and redshift (right panel). Each row includes $\sim$70 emission-line selected galaxies. The best-fit continuum is subtracted from each spectrum, and the spectra are normalized by the inverse of their luminosity distance squared---so that the brightness corresponds to the luminosity of the sources. Prominent emission line/line complexes in the optical-NIR are indicated.}
\end{figure*}



\begin{deluxetable*}{rcl}
\tablecolumns{3}
\tablewidth{0pt}
\tablecaption{Description of the Eazy-py Catalog\label{table:eazy_catalog}}
\tablehead{
\colhead{column} & 
\colhead{units}  &
\colhead{description}}
\startdata
id & & Galaxy ID, matched to \citet{skelton14}\\
ra & [deg] & Right Ascension, J2000\\
dec & [deg] & Declination, J2000\\
z\_spec & &  Ground-based spectroscopic redshift (if available)\\
nusefilt & &Number of filters used for photo-z\\
lc\_min &[\AA] & Minimum effective wavelength of valid filters\\
lc\_max &[\AA] & Maximum effective wavelength of valid filters\\
z\_phot & & Photometric Redshift, maximum likelihood\\
z\_phot\_chi2 & & $\chi^2$ at z\_phot\\
z\_phot\_risk & & Risk evaluated at z\_phot\\
z\_min\_risk & & Redshift where risk is minimized\\
min\_risk & & Minimized risk\\
z\_raw\_chi2 & & Redshift at the minimum $\chi^2$\\
raw\_chi2 & & Minimum $\chi^2$\\
z025, z160, z500, z840, z975 & & Confidence intervals of redshift\\
rest{\tt{[FILT]}} & & Rest-Frame flux in {\tt{[FILT]}}-band\\
rest{\tt{[FILT]}}\_err & & Uncertainty of ``rest{\tt{[FILT]}}''\\
dL &  [Mpc] & Luminosity distance at z\_phot\\
mass & [M$_{\odot}$] & Stellar mass \\
sfr & [M$_{\odot}\,\mathrm{yr}^{-1}$] & Star-Formation Rate\\
Lv &  [L$_{\odot}$] & V-band Luminosity\\
LIR & [L$_{\odot}$] & Total 8-1000 $\mu$m luminosity\\
MLv &  [M$_{\odot}$/L$_{\odot}$] & Mass-to-light ratio in V-band\\
Av & [mag] & Extinction in V-band\\
mass\_p  & [M$_{\odot}$] & Confidence intervals of mass\\
sfr\_p  & [M$_{\odot}\,\mathrm{yr}^{-1}$] & Confidence intervals of sfr\\
Lv\_p   & [L$_{\odot}$] &Confidence intervals of Lv\\
LIR\_p  & [L$_{\odot}$] & Confidence intervals of LIR\\
ssfr\_p & [yr$^{-1}$] & Confidence intervals of specific star-formation rate\\
DISTMOD &  & Distance Modulus\\
ABSM\_271 & [mag] & Absolute Magnitude at 1700 \AA\\
ABSM\_272 & [mag] &  Absolute Magnitude at 2200 \AA\\
ABSM\_274   &[mag] &  Absolute Magnitude at 2800 \AA\\
  \enddata
        \tablenotetext{a}{see the Eazy-Py documentation (\url{eazy-py.readthedocs.io}) for full details.}
        \tablenotetext{b}{{\tt[FILT]} is the filter.}
        \tablenotetext{c}{``risk" is defined in \citet{2018PASJ...70S...9T}.}
\end{deluxetable*}

\subsection{Constraints on the Lyman-$\alpha$ Equivalent Width Distribution Function}

A main goal of CLEAR is to constrain the \lya\ emission from galaxy candidates at $z > 6$.   There are several advantages in the use of space-based, slitless spectroscopy to explore \lya\ in galaxies, and these are related to the fact that the \hst/WFC3 grism data provide independent observations of the emission in these galaxies without many of the biases, backgrounds, and selection effects that can impact ground-based studies.   These include the following.   

First, space-based slitless spectroscopy eliminates systematic differences caused by using different optical / near-IR spectrographs from ground-based telescopes.  Typically optical spectrographs are used to study \lya\ at $z < 7$, and near-IR spectrographs are used for higher redshifts (see, e.g., \citealt{jung2018,jung2019, Pentericci18, Hoag19, Fuller20}).

Second, ground-based surveys have different instrumental sensitivities, seeing variations, and variable night-sky line emission.   This produces a time and wavelength-dependent flux sensitivity to emission lines with variations of a factor of $>$5 (e.g., \citealt{treu2012,jung2020}).  Furthermore, there is evidence that emission line fluxes from ground-based spectra are 2--4$\times$ lower than slitless, space-based data, which can result from (seeing-dependent) slit-loss corrections and difficulties in flux calibrations in the case that only an emission line is detected with no continuum (e.g., see \citealt{masters2014}).  In contrast, space-based observations provide a smoothly varying line-flux sensitivity function (e.g., \citealt{jung2021}).  

Third, slitless spectroscopy targets galaxies indiscriminately---with no target pre-selection.  A single \hst/WFC3 G102 observation is sensitive to \lya\ from all galaxies with $6.0 < z < 8.2$ (see Figure~\ref{fig:lines_redshift}).  Ground-based observations require slits and potentially imperfect/biased selection. For instance, there is a natural bias to place slits on brighter objects (and these often have the lower \lya\ EW, see \citealt{stark2010,finkelstein2013,oesch2015,jung2018,jung2020}).   Slitless, space-based spectra therefore provide less biased samples compared to ground-based spectra in this regard.   

In reality, there are few objects with plausible detections of \lya\ in galaxies at $z > 6$ in the \clearer{} dataset.  \citet{jung2021} identify several possible candidates, including one galaxy at $z=6.51$.   This (lack of) strong detections is consistent with other searches for \lya\ from \hst/WFC3 grism data, where there are only a handful of detections in galaxies at $6 < z < 8$ \citep[e.g.,][]{schmidt16,tilvi16,larson18,jung2021}.  The basic conclusion is that the \lya\ emission from all galaxies is substantially weaker than in lower-redshift galaxies (whose \lya\ EW distributions were used to predict \lya\ line fluxes at $z > 7$).   This is particularly true for galaxies with flower UV luminosities, as these are expected to be strong \lya\ emitters. 

Using CLEAR we have obtained improved constraints on the evolution of the \lya\ EW distribution function.  In \citet{jung2021}, we combined observations from CLEAR with ground-based datasets to measure this evolution.  We found that for all galaxies at $z > 6$, \lya\ emission is significantly suppressed compared to samples at $z < 6$.  Interestingly, however, \citeauthor{jung2021} argue there is tentative evidence that the suppression of \lya\ is \textit{stronger} for galaxies with lower UV luminosities. This means that there is additional attenuation/absorption of \lya\ photons in lower luminosity galaxies.  This can be explained if reionization is highly inhomogeneous, where the more UV luminous galaxies blow larger ionized ``bubbles'' around them (e.g., \citealt{Finlator09, Pentericci14, Katz19}).  Once these bubbles reach sizes of $\sim$1 physical Mpc, then the \lya\ photons from the source have been sufficiently redshifted (compared to the Hubble flow) that attenuation is mitigated (e.g., \citealt{Mason20, Park21, Qin22, Smith22}).  In this way, the \lya\ photons from UV brighter objects are less impacted than lower-luminosity objects that have small ionized bubbles surrounding them.   However, as we discuss in \citeauthor{jung2021}, the constraints based on the current datasets are still too small given the sample size, but this makes predictions for both \textit{JWST} and \textit{NGRST} that should identify \lya\ emission at fainter flux sensitivities and for vastly larger samples. 

\subsection{Studies of Stellar Populations in Distant Galaxies}

Several studies have studied the stellar populations of galaxies as derived from their stellar continuum features in the \clearer{} dataset.  

For high-redshift galaxies, the G102 and G141 grism data probe many of the well known spectral features of stellar populations in the rest-frame optical. Due to uncertain and time-variable sky backgrounds, these features are difficult (but not impossible) to detect in high-redshift galaxies from the ground. Figure~\ref{fig:quiescent_example} shows a stacked spectrum of galaxies selected to be ``quiescent'' based on their $UVJ$ rest-frame colors. These are a subset of those galaxies studied by \citet{estrada-carpenter19, estrada-carpenter20}.   The shape of the spectrum and strength of spectral features are sensitive to age and metallicity (see also, discussion in \citealt{estrada-carpenter19,estrada-carpenter20}).  The spectrum on the blue side is consistent with Solar metallicity.  The shape of the spectrum on the red side is more consistent with super-Solar abundances.  These facts could be an indication of higher $\alpha$/Fe ratios (such as titanium, magnesium, oxygen) at fixed [Fe/H], consistent with other observations of quiescent galaxies at low and high redshifts \citep{Conroy_2012,Choi_2014,Kriek_2019}.  This could also be related to age as the galaxies in the stack span a range of redshift, with higher redshift galaxies contributing more to the rest-frame blue wavelengths.  At higher-redshift, quiescent galaxies show evidence of younger stellar populations \citep[e.g.,][]{estrada-carpenter19} and so the differences in the spectra could be representative of differences in population age. 

\citet{estrada-carpenter19} used the G102 data for a sample of quiescent-selected galaxies at $1 < z < 1.8$ to measure constraints on the ages, star-formation histories, and metallicities of the galaxies.  They showed that massive quiescent galaxies at these redshifts already harbor older stellar populations (with light-weighted ages indicating formation epochs $z_\mathrm{f} > 2.5$) and that they had already enriched to stellar metallicities approaching or exceeding Solar ($\approx Z_\odot$).  This indicates that the massive $z>1$ quiescent galaxies experienced very early star-formation and chemical enrichment.   \citet{estrada-carpenter20} used models with flexible star-formation histories to show that quiescent galaxies with the earliest formation have more compact morphologies, indicating a correlation between formation and compactness. The stellar population constraints on stellar mass, dust attenuation and SFRs have also been used to study galaxy gas-phase metallicity--mass and ionization--mass relations \citep{papovich22} (see next subsection) and to study the evolution of ``green valley'' galaxies in transition from the star-forming to quiescent sequences (Estrada-Carpenter et al., in prep).

One of the big remaining questions is, where are the progenitors of the massive, quiescent galaxies with Solar enrichment?   Measurements of the \textit{stellar} metallicites (derived from continuum spectroscopy, including our own work with CLEAR) of massive, quiescent galaxies at $z \gtrsim 1$ are consistent with Solar abundances \citep[e.g.,][]{onodera15,kriek16,Kriek_2019,estrada-carpenter19,Lonoce_2020}. Most studies of the gas-phase metallicities of star-forming galaxies  at $z \simeq 1-2$ (including our own analysis from CLEAR) find that star-forming galaxies with stellar mass $\log M_\ast/M_\odot \gtrsim 10.5$ are sub-solar \citep[e.g.,][]{steidel14,sanders15,strom17,henry21,papovich22}, with $12 + \log(\mathrm{O/H})$ approximately 0.2--0.3 dex below the Solar value ($12+\log(\mathrm{O/H})_\odot = 8.69$, \citealt{asplund09}).   Therefore, we are missing those star-forming galaxies at $z \gtrsim 2$ that have high-levels of enrichment, similar to the metallicities inferred for quiescent galaxies at this epoch.  This discrepancy is compounded by evidence that the $\alpha$/Fe ratios are enhanced \citep{steidl16,strom18,strom22, topping20}, which requires even lower iron abundances (which typically dominate $Z_\ast$, see \citealt{estr19}).    Oxygen (an $\alpha$ element) stems primarily from observations of the nebular gas, while the iron abundance dictates the shape of the stellar continuum, and direct measures of $\alpha$/Fe from stellar continuum are only possible for rare cases of bright galaxies, see, e.g., \citet{kriek16,Kriek_2019}.  As illustrated in Figure~\ref{fig:quiescent_example}, the shape of the spectrum and the strength of the spectral features are sensitive to changes in the metallicity.  This presents an opportunity for observations from future studies with \jwst\ and \ngrst\ that will enable both higher-quality data and larger samples of galaxies to explore this problem of the ``missing'', high-metallicity progenitors of the quiescent galaxies at $z \gtrsim 1$.

\subsection{Emission Line Diagnostics of Galaxies} 

The \clearer{} dataset enables a large range of science using galaxy emission lines as diagnostics of star-formation, gas conditions including ionization and metallicity, and black hole activity.

In Figure \ref{fig:stack_heatmap}, we show stacks of the 1D \clearer{} spectra in bins of stellar mass and redshift. Each row in the stack contains $\sim$70 individual galaxy spectra. The stacks illustrate the change in the relative luminosity of the emission lines as a function of stellar mass (left panel) and redshift (right panel), and indicate which strong rest-optical lines are accessible (although not necessarily detected in individual galaxies) in the \clearer{} spectral range.


Several papers have used the \clearer{} dataset to investigate how star-formation proceeds in galaxies from the local Universe to cosmic high noon \citep{cleri22a, matharu22}.  \citet{cleri22a} identified galaxies at $z < 0.29$ where both H$\alpha$ and Pa$\beta$ emission are detected in the integrated (1D) CLEAR spectra.  Because these are both hydrogen recombination lines, their theoretical line ratio is relatively insensitive to conditions in the \ion{H}{2} regions (over a wide range of temperature and density, see \citealt{2006agna.book.....O}).   By comparing the Pa$\beta$ emission to dust-corrected UV emission, one can probe star-formation on $\sim$5 Myr timescales (the lifetimes of O-type stars responsible for ionizing the nebula which produce emission from hydrogen recombination) with $\sim$100 Myr timescales (the lifetimes of B-type stars responsible for the UV continua). \citet{cleri22a} showed that low-mass galaxies have more scatter in the Pa$\beta$/UV ratios than higher-mass galaxies, and they argued that this is a result of increased burstiness in the galaxies' star-formation histories.

\citet{matharu22} used the spatially resolved H$\alpha$ emission for galaxies in \clearer{} to study how star-formation proceeds at $0.5 < z < 1$.  They showed that the sizes of H$\alpha$ disks are larger than that of the stellar continuum (measured in a broadband that covers the same rest-frame wavelengths as the H$\alpha$ emission), but that there is redshift evolution compared to samples at $1 < z < 1.5$ (from 3DHST, \citealt{nelson16}) and $z\sim 2$ (from KMOS$^{3D}$, \citealt{wilman20}). \citeauthor{matharu22} showed that this evolution is consistent with star-formation proceeding in an ``inside--out'' fashion in galaxies.

The \clearer{} data have also been used to study the physical conditions of the nebular \citep{simons21, papovich22, backhaus22a, backhaus22b} and highly-ionized \citep{cleri22b, cleri2023} gas in galaxies.  

\citet{papovich22} used measurements of the fluxes (and flux ratios) of the \oii, \oiii, and H$\beta$ emission lines from the CLEAR spectra to study the ionization and chemical enrichment of galaxies over $1.1\,<\,z\,<\,2.3$. They showed that at fixed stellar mass ($\log$ M$_*$/M$_{\odot}\,\sim9.4-9.8$), higher-redshift galaxies have lower gas-phase metallicities and higher ionization parameters than they do at lower redshift ($z\,\sim\,0.2$). Moreover, at fixed mass and/or at fixed metallicity, higher-redshift galaxies have ionization parameters that correlate positively with their specific star-formation rate (sSFR). \citeauthor{papovich22} posit that this correlation could arise because the gas density (and/or gas geometry) and the escape fraction of ionizing photons likely both increase with increasing sSFR.

\citet{backhaus22a} used the CLEAR spectra to study the spatially-integrated line ratios \oiii/H$\beta$, \sii/(H$\alpha$+\nii), and \neiii/\oii{} in galaxies spanning $0.6\,<\,z\,<\,2.5$. Comparing with photoionization models, they conclude similarly to \citet{papovich22} that galaxies with higher mass and lower star-formation rates have both higher metallicity and lower ionization parameters. To distinguish ionization from star-formation and AGN, they construct the diagnostic diagrams `OHNO' (\oiii/H$\beta$ vs \neiii/\oii) and `unV087' (\oiii/H$\beta$ vs \sii/(H$\alpha$+\nii)) of this sample. The `un' in `unV087' indicates that H$\alpha$ + \nii{} are spectrally-unresolved in the \hst{} grisms. While the `unV087' diagram poorly separates the AGN and [NeV]-emitting galaxies (indicating high ionization) from the star-forming galaxies, \citeauthor{backhaus22a} argue that the `OHNO' diagram does effectively discriminate these populations. They conclude that `OHNO' will be a useful indicator for AGN activity and the ionization conditions in high-redshift galaxies observed by the \emph{JWST} observatory.

\citet{backhaus22b} measured the radial gradient of the \oiii/H$\beta$ ratio in galaxies over $0.6\,<\,z\,<\,1.3$ to study the spatial variations of their ionization and to search for low-luminosity AGN. While the majority of the galaxies are consistent with a zero gradient, they argue that 6-16\% of the galaxies in the sample likely have nuclear \oiii/H$\beta$ ratios that are 0.5 dex higher than they are in their outer regions. \citeauthor{backhaus22b} argue that these galaxies (which are generally not detected in X-rays) may host low-luminosity AGN. Furthermore, they did not find evidence for a significant population of sources with off-nuclear ionization. 

\citet{simons21} used the spatially-resolved maps of the \oii, H$\beta$, \oiii, H$\alpha$+\nii, and \sii{} emission lines from \clearer{} to derive gas-phase metallicity maps for 238 star-forming galaxies over the redshift range $0.6\,<\,z\,<\,2.6$. They measured the radial gradient of these metallicity maps, and report that the majority of galaxies at this redshift are consistent with a null or positive (aka ``inverted'', i.e., more metal-rich in galaxy outskirts) metallicity gradient (see also e.g., \citealt{wang17, wang19, wang20, Curti20, Li22}). This finding is somewhat puzzling because it runs counter to simple expectations from star-formation and stellar evolution. In star-forming galaxies at this redshift, the star-formation surface density is on average higher in the galaxy centers than it is in the galaxy outskirts \citep{nelson16, Tacchella18}. Given that, we might expect for these galaxies to gradually form negative metallicity gradients (more metal-rich in the galaxy centers) through stellar evolution and local chemical enrichment. \citet{simons21} argue that the ubiquity of null/positive gradients in these galaxies implies that their gas-phase metals are being re-distributed on galaxy scales (or their ISM is being unevenly diluted through metal-poor accretion) on timescales shorter than the short time ($\lesssim$100 Myr, \citealt{simons21}) it would take for them to naturally develop a declining metallicity gradient through stellar evolution. These metals could be re-distributed around galaxies through galactic scale outflows and/or the high levels of turbulence in the interstellar medium of these galaxies (e.g., \citealt{weiner06, FS06, FS09, kassin07, kassin12, wisnioski15, simons16, simons17, ubler19, price20}).

Finally, \citet{cleri22b} used the CLEAR 1D spectra to search for the high-ionization [NeV] ($\lambda$3426) emission in galaxies over $1.4\,<\,z\,<\,2.3$. [NeV] has an exceedingly high creation potential (97.11 eV), and is an indicator for highly-energetic photoionization---e.g., from AGN, SNe radiation, and/or a hard ionization spectrum from stars (Cleri et al., in prep). \citet{cleri22b} select 25 galaxies in the CLEAR sample with [NeV] detected. Based on the ratios of the \oiii/H$\beta$ lines in these galaxies, they show that most of the sample is consistent with photoionization from an AGN. 

\section{Summary}\label{section:summary}

This paper presents an overview of the CANDELS Lyman-$\alpha$ Emission at Reionization (CLEAR) survey---a 130 orbit \textit{Hubble Space Telescope}/Wide Field Camera 3 (\hst{}/WFC3) spectroscopic and imaging program. The CLEAR observations include 10- and 12-orbit \hst{}/WFC3 G102 grism spectroscopy and F105W imaging in the GOODS-N and GOODS-S legacy fields, respectively. The full dataset discussed here includes the WFC3/G102 grism observations from the CLEAR survey and overlapping WFC3/G102 + G141 observations from a number of ancillary programs in the \hst{} archive. 

We discuss the design of the CLEAR survey, the data processing and products, and the science that has been carried out by the CLEAR team with this dataset. Alongside this \textit{Paper}, we release a number of science-ready data products created from this program, including: emission line flux catalogs, updated 3D-HST photometric catalogs, and 2D and 1D extracted spectra for \numextracted{} galaxies. These products are available at MAST\footnote{\url{https://archive.stsci.edu/hlsp/clear/}} as a High Level Science Product via \dataset[10.17909/9cjs-wy94]{\doi{10.17909/9cjs-wy94}}.

\section*{Acknowledgements}

We thank the anonymous referee and data editor for a constructive report that improved this manuscript. We thank Mark Dickinson, Rachael Livermore, and Ryan Quadri for valuable conversations and contributions to the early development of the CLEAR survey. This work is based on data obtained from the Hubble Space Telescope through program number GO-14227.  Support for Program number GO-14227 was provided by NASA through a grant from the Space Telescope Science Institute, which is operated by the Association of Universities for Research in Astronomy, Incorporated, under NASA contract NAS5-26555.  This work is supported in part by the National Science Foundation through grant AST 1614668. RCS appreciates support from a Giacconi Fellowship at the Space Telescope Science Institute. VEC acknowledges support from the NASA Headquarters under the Future Investigators in NASA Earth and Space Science and Technology (FINESST) award 19-ASTRO19-0122, as well as support from the Hagler Institute for Advanced Study at Texas A$\&$M University. Portions of this research were conducted with the advanced computing resources provided by Texas A$\&$M High Performance Research Computing.  This work was supported in part by NASA contract  NNG16PJ33C, the Studying Cosmic Dawn with WFIRST Science Investigation Team. This work benefited from generous support from the George P. and Cynthia Woods Mitchell Institute for Fundamental Physics and Astronomy at Texas A$\&$M University.   CP thanks Marsha L.\ and Ralph F.\ Shilling for generous support of this research.   This research made use of Astropy (http://www.astropy.org) a community-developed core Python package for Astronomy \citep{astropy:2013, astropy:2018}. Some/all of the data presented in this paper were obtained from the Mikulski Archive for Space Telescopes (MAST) at the Space Telescope Science Institute. The specific observations analyzed can be accessed via \dataset[10.17909/9cjs-wy94]{\doi{10.17909/9cjs-wy94}}. 

\software{eazy-Py \citep{Brammer21}, EAZY \citep{brammer08}, Grizli \citep{2019ascl.soft05001B}, AstroDrizzle \citep{2012drzp.book.....G}, Astropy \citep{astropy:2013, astropy:2018}, NumPy \citep{harris2020array}, Matplotlib \citep{Hunter2007}}

\bibliography{clearsurvey}{}

\end{document}